%% file: ms.tex
\documentclass[format=acmsmall, review=false, screen=false, acmcopyrightmode=none]{acmart}

\usepackage{booktabs} 
\usepackage{paralist}
\usepackage{ulem}

\usepackage[ruled]{algorithm2e} 

\SetAlFnt{\small}
\SetAlCapFnt{\small}
\SetAlCapNameFnt{\small}
\SetAlCapHSkip{0pt}
\IncMargin{-\parindent}

\acmJournal{CSUR}
\acmVolume{99}
\acmNumber{99}
\acmArticle{00}
\acmYear{0000}
\acmMonth{9}
\copyrightyear{0000}

\setcopyright{acmlicensed}

\acmDOI{0000001.0000001}

\received{September 2017}
\received[revised]{XXXX XXXX}
\received[accepted]{XXXX XXXX}

\usepackage{xcolor}

\usepackage{amssymb}
\usepackage{multirow}
\usepackage{lscape}

\begin{document}
\title[A Survey of Insider Threat Taxonomies, Analysis, Modeling, and Countermeasures]{Insight into Insiders and IT: A Survey of Insider Threat Taxonomies, Analysis, Modeling, and Countermeasures}

	\author{Ivan Homoliak}		
\affiliation{%
	\institution{STE-SUTD Cyber Security Laboratory, Singapore University of Technology and Design}
	\streetaddress{8 Somapah Road}
	\postcode{487372}
	\country{Singapore}}
\affiliation{%
	\institution{Faculty of Information Technology, Brno University of Technology}
	\streetaddress{Bozetechova 1/2, Brno}
	\postcode{61266}
	\country{Czech Republic}}

\author{Flavio Toffalini}
\author{Juan Guarnizo}
\affiliation{%
	\institution{STE-SUTD Cyber Security Laboratory, Singapore University of Technology and Design}
	\streetaddress{8 Somapah Road}
	\postcode{487372}
	\country{Singapore}}

\author{Yuval Elovici}
\affiliation{%
	\institution{STE-SUTD Cyber Security Laboratory, Singapore University of Technology and Design}
	\streetaddress{}
	\postcode{}
	\country{}}

\author{Mart\'{i}n Ochoa}
\affiliation{%
	\institution{STE-SUTD Cyber Security Laboratory, Singapore University of Technology and Design}
	\streetaddress{8 Somapah Road}
	\postcode{487372}
	\country{Singapore}}
\affiliation{%
	\institution{Department of Applied Mathematics and Computer Science, Universidad del Rosario}
	\streetaddress{Calle 12c \#6-25, Bogot\'{a}}
	\postcode{}
	\country{Colombia}}

\begin{abstract}
Insider threats are one of today's most challenging cybersecurity issues that are not well addressed by commonly employed security solutions.
Despite several scientific works published in this domain, we argue that the field can benefit from our proposed structural taxonomy and novel categorization of research that contribute to the organization and disambiguation of insider threat incidents and the defense solutions used against them.
%
%
The objective of our categorization is to systematize knowledge in insider threat research, while using existing grounded theory method for rigorous literature review. 
The proposed categorization depicts the workflow among particular categories that include: 1) \textit{Incidents and datasets}, 2) \textit{Analysis of incidents}, 3) \textit{Simulations}, and 4) \textit{Defense solutions}.
Special attention is paid to the definitions and taxonomies of the insider threat; we present a structural taxonomy of insider threat incidents, which is based on existing taxonomies and the 5W1H questions of the information gathering problem.
Our survey will enhance researchers' efforts in the domain of insider threat, because it provides: a) a novel structural taxonomy that contributes to orthogonal classification of incidents and defining the scope of defense solutions employed against them, b) an overview on publicly available datasets that can be used to test new detection solutions against other works, c) references of existing case studies and frameworks modeling insiders' behaviors for the purpose of reviewing defense solutions or extending their coverage, and d) a discussion of existing trends and further research directions that can be used for reasoning in the insider threat domain.
\end{abstract}

%
%


%
%



\settopmatter{printacmref=false, printccs=false, printfolios=true}

\setcopyright{none}

\maketitle

\renewcommand{\shortauthors}{I. Homoliak et al.}

\input{body-CSUR}
\end{document}

%% file: body-CSUR.tex
\citestyle{acmauthoryear}
\setcitestyle{square}

\section{Introduction}

Insider threats are one of the most challenging attack models to deal with in practice. 
According to a recent survey, 27\% of all cyber crime incidents were suspected to be committed by insiders, and 30\% of respondents indicated that the damage inflicted by insiders was more severe than the damage caused by outside attackers~\cite{2017-Trzeciak}.
Similar numbers are reported by~\cite{2016-Collins-CERTv5}: ``23\% of electronic crime events were suspected or known to be caused by insiders,'' while 45\% of the respondents thought that the consequences were worse than the consequences of outsiders.
In another survey investigating economic crime~\cite{2017-PwC-Economy-Crime}, internal fraudsters acted as the main perpetrator in 29\% of cases.
According to a survey conducted by Vormetric, Inc.~\cite{2015-Vormetric}, only 11\% of respondents felt that their organization was not vulnerable to insider attacks, while 89\% felt at least somewhat vulnerable to insider attacks.

In recent years, famous whistle-blowers have made their way into the media, such as the high profile data leakage cases involving Edward Snowden or Chelsea Manning (see~\cite{2016-Collins-CERTv5} for a collection of famous insider threat cases).
Although such cases may be viewed as security issues breaking the confidentiality of secret information, they may also be viewed as human loyalty manifested to the country or the society (see Section~\ref{sec:Taxonomies} for further discussion). 

In general, insiders are authorized users that have legitimate access to sensitive/confidential material, and they may know the vulnerabilities of the deployed systems and business processes.
Many attacks caused by malicious insiders are more difficult to detect compared to those of external attackers whose footprints are harder to hide~\cite{2016-Moore}.
In addition, there has been an increasing trend of unintentional insider threat in recent years~\cite{2016-Collins-CERTv5}. 
Therefore, the motivation for dealing with insider threat is very high and is likely to grow. 

Concerns about insider threat in the literature are not new, and there is an impressive body of knowledge in this broad field.
In the last decade there have also been several attempts to survey this field.
However, after reviewing such works, we encountered various shortcomings and identified the need for an up-to-date, more comprehensive survey.
For instance, some surveys focus exclusively on detection approaches~\cite{2008-Salem,2016-Gheyas,2009-Bertacchini,2016-Sanzgiri} or lack a systematical approach to the categorization of the literature~\cite{2014-Azaria}.
The objective of this work is thus to conduct an extensive literature review in the field of insider threat, while aiming at systematization of the knowledge and research conducted in this area.
We view this as important for both the researchers that design experimental defense solutions, as well as the security practitioners who seek to understand the problem and are tasked with selecting or implementing appropriate defense solutions for their specific needs.

\subsection{\textbf{Survey Approach}} \label{sec:SurveyApproach}
Our main objective for this work is to address the identified gaps and consolidate the information contained in existing surveys, by including a more exhaustive and up-to-date literature set, emphasizing the review and unification of taxonomies, and using a systematical approach for the categorization of the literature.
To this end, we have revised and updated the bibliography of previous surveys, and to ensure that we have not inadvertently left out any relevant work, we manually checked the top 100 best ranked papers in the field according to Google Scholar, as well as the 100 most cited papers in the Web of Science databases, by querying the term \textit{insider threat}.
Overall, our sample set contained 322 works, and 108 of them were filtered out based on our inclusion and exclusion criteria (explained below). 
Note that given the vast amount of work in the field, the goal of this work was not to exhaustively cover all of the literature, but to select a large enough literature set to review the state-of-the-art and to propose a reasonable categorization of it.

\paragraph{\textbf{Survey Scope.}} 
The scope of our survey is based on the following criteria:
\textbf{a)} The articles included in this survey were selected based on a widespread view of the insider threat problem, ranging from definitions and taxonomies of insider threat, and analysis and modeling of incidents, to conceptual defense solutions and their proofs of concept;
\textbf{b)} We focused on studies for which the insider threat problem was the main subject; therefore, we excluded papers that only mentioned the insider threat problem tangentially.
We did, however, include several examples of masquerade detection approaches, which are related to the identity theft problem but are also considered part of the insider threat problem;
\textbf{c)}  We excluded non-unique papers from the same group of authors who presented a concept/approach across multiple papers. 
The older version of a study was usually superseded by the newer one, except in cases in which the newer version contained fewer details or also focused on another idea.		

After specifying the scope of the survey, we applied an iterative process to construct a literature categorization based on grounded theory for rigorous literature review~\cite{wolfswinkel2013using}, which serves as guidelines on an analysis and presentation of findings in a particular field of research.
While processing the set of papers, we identified several abstract concepts, proposed a workflow-based categorization, and rearranged the bibliography according to it.
The proposed categorization consists of a review of datasets, case studies, analysis and modeling of incidents, simulations, conceptual and practical defense solutions, and best practices.
Note that our main categories are not meant to be disjoint, as there are works that address aspects of various proposed categories.
However, we believe that our categorization offers useful dimensions with which to classify works in the literature, while enabling researchers to identify relevant related work.

\subsection{Contributions}
In sum, this work presents a novel insider threat survey that aims to be comprehensive, yet succinct and easy to access by researchers looking for new avenues to explore or to learn about the subject.
The main contributions of this survey can be summarized as follows:
\textbf{a)} To the best of our knowledge, this is the first work that systematically categorizes heterogeneous insider threat studies and thereby enables readers to obtain a panoptic view on this disparate topic. 
\textbf{b)} We survey existing taxonomies of the insider threat problem, and based on them, we propose a practical and unified taxonomy that can be used to classify: 1) an insider threat incident, or 2) specialization/coverage of a defense solution.
\textbf{c)} We aggregate information about publicly available datasets that can be utilized for testing a detection solution and comparison with other works included in this survey that have previously used the datasets.
\textbf{d)} We identify open questions and challenges in insider threat detection, prevention, and mitigation, while proposing further research directions.

The rest of the paper is organized as follows.
Section~\ref{Sec:Existing-Surveys} briefly reviews existing surveys in the field and also mentions their contributions, shortcomings, and how our work differs from them.
In Section~\ref{sec:Taxonomies} we discuss the scope of insider threat and provide a survey of definitions and existing taxonomies; based on existing taxonomies, we propose a structural taxonomy of insider threat incidents.
After providing this background, we change our focus to the categorization of research conducted in this domain, which is discussed in Section~\ref{Sec:Proposed-Categorization}.
A detailed description and the subcategorization of specific major categories is presented as follows: incidents and datasets are dealt within Section~\ref{Sec:Incidents}, an analysis of incidents is provided in Section~\ref{Sec:Analysis-of-Incidents}, simulation research is presented in Section~\ref{Sec:Simulations}, and defense solutions are covered in Section~\ref{Sec:Defense}.
The last part -- Section~\ref{Sec:Conclusion} -- concludes the paper and proposes further directions in the field.

\section{Existing Surveys}\label{Sec:Existing-Surveys}
In this section we provide a brief summary of existing surveys involving insider threats, and then we describe how our proposed taxonomy differs from them.
Salem et al. ~\citeyear{2008-Salem} introduced a taxonomy of malicious insiders, dividing them into two categories according to the knowledge they have about the target system: traitors and masqueraders.
The authors reviewed the literature on insider detection works and divided the works into three types of approaches: a) ``host-based user profiling approaches,'' b) ``network level approaches,'' and c) ``integrated approaches.''
Network level and host-based profiling may have, according to the authors, a high chance of detecting traitors, while host-based user profiling may be successful in identifying masqueraders.
Also, the authors posited that malicious actions of insiders occur at the application and business process levels.
Hunker and Probst~\citeyear{2011-Hunker} proposed a categorization of the research based on a combination of psycho-social input data with technical data. 
The resulting categories consist of three types of approaches to insider threat detection: 1) ``sociological, psychological, organizational,'' 2) ``socio-technical,'' and 3) ``technical.'' 
The authors emphasized that successful insider threat detection techniques require a combination of various approaches.
Azaria et al.~\citeyear{2014-Azaria} divided related works into six categories, likely based on the most significant trends in the field: 1) ``psychological and social theories,'' 2) ``anomaly-based approaches,'' 3) ``honeypot-based approaches,'' 4) ``graph-based approaches,'' 5) ``game theory approaches,'' and 6) ``motivating studies.'' 
However each category draws from a different dimension of possible categorizations, and thus a new approach may not fit any of the proposed categories.
The authors aimed to reinterpret principles and results of the approaches included in their survey. 
Ophoff et al.~\citeyear{2014-Ophoff} conducted a literature review and classified insider threat research based on ``50 top ranked MIS journals from the ISI world, as well as top security journals in the IS domain.'' 
The search term \textit{``insider threat"} yielded over 600 results, however after removing irrelevant research and filtering duplicates they had 90 papers, which they divided into six categories and 13 subcategories using the grounded theory approach~\cite{wolfswinkel2013using}. 
The categorization is composed of: 1) ``insider threat mitigation,'' 2) ``theoretical perspectives,'' 3) ``insider threat management,'' 4) ``insider threat behavior,'' 5) ``insider threat overview'', and 6) ``miscellaneous.'' 
This work focused on categorization, rather than focusing on providing an overview of relevant papers in each category.
Gheyas and Abdallah~\citeyear{2016-Gheyas} applied ``a systematic literature review and meta-analysis'' defined by PRISMA~\cite{moher2009preferred} for a review of malicious insider threat detection approaches and their concepts. 
The authors ranked research papers considering the theoretical properties and the clarity of the contributions presented. 
They also proposed a categorization of studies according to the input dataset, features used, detection strategy implemented, and the underlying machine learning algorithm. 
The authors discussed key challenges in malicious insider threat detection from the big data perspective, stated trends in the field, and provided best practice recommendations for future research.
Sanzgiri and Dasgupta~\citeyear{2016-Sanzgiri} proposed classification of insider threat detection methods based on strategies and features used in the detection itself, introducing the following nine classes: 1) ``anomaly-based,'' 2) ``role-based access control,'' 3) ``scenario-based,'' 4) using ``decoys and honeypots,'' 5) ``risk analysis using psychological factors,'' 6) ``risk analysis using workflow,'' 7) ``improving defense of the network,'' 8) ``improving defense by access control,'' and 9) ``process control to dissuade insiders.''
Bertacchini and Fierens~\citeyear{2009-Bertacchini} conducted a literature review on masquerader detection approaches in the Unix command domain and divided them into several categories: 1) ``information-theoretic-based,'' 2) ``text mining-based,'' 3) ``HMM-based,'' 4) ``Na\"{i}ve Bayes-based,'' 5) ``sequence-based and bioinformatics-based,'' 6) ``SVM-based,'' and 7) ``other approaches.'' 
In addition, the authors summarized the properties and the Unix command datasets used in specific works.

\subsection{Comparison with Our Survey}
In sum, the existing surveys deal with either categorization of \textit{heterogeneous studies}~\cite{2014-Azaria,2014-Ophoff,2011-Hunker} or \textit{homogeneous studies} that are constrained to detection approaches~\cite{2008-Salem,2016-Gheyas,2009-Bertacchini,2016-Sanzgiri}.
A characteristic of homogeneous studies is that they are limited to specific application domains: masquerader detection approaches~\cite{2009-Bertacchini}, insider threat detection approaches~\cite{2016-Gheyas,2016-Sanzgiri}, and host-based/network-based profiling of insiders~\cite{2008-Salem}.
Since we deal with a broad scope of research related to insider threat, our survey involves heterogeneous studies, and thus, is more similar to~\cite{2014-Azaria,2014-Ophoff,2011-Hunker}.
However, in contrast to existing surveys, we first perform coarse-grained categorization based on the workflow of research efforts, and then we make fine-grained categorization within each category.

\section{Definitions and Taxonomies}\label{sec:Taxonomies}
This section focuses on definitions and taxonomies of insider threats. 
First, we start by emphasizing the difficulty of determining the border of distinction for insider threats and also intent of whistle-blowers. 
Then, we survey several disparate definitions of insiders and insider threats, followed by brief descriptions of state-of-the-art taxonomies, which we split into three types according to an insider's intention: \textit{malicious}, \textit{unintentional}, and special case of taxonomies involving both types of intentions, denoted as \textit{combined} taxonomies.
Afterwards, based on the taxonomies described, we propose a structural taxonomy of insider incidents by providing answers to questions regarding the information gathering problem.
\vspace{-0.1cm}
\paragraph{\textbf{Scope of Insider Threat}}
An inherent problem associated with defining the scope of insider threats is that often it is difficult to distinguish between insiders and outsiders of an organization once they are operating within an internal network. 
Further complicating matters is the fact that there are insiders that \textit{attack from outside.} 
One representative example of such an insider is an ex-employee who recently left the organization.

According to Neumann~\citeyear{2010-Neumann}, outsiders who have successfully penetrated into an IS or network are considered outsiders unless they have obtained enough knowledge to make them indistinguishable from the insiders they are masquerading as; the author considers a Turing test to make such a differentiation. 
Moreover, insider threat has also been defined as part of broader topics that have emerged throughout the history of computer systems, such as intrusion attempts~\cite{1980-Anderson}, threats to IS security~\cite{1992-Loch}, and counterproductive workplace behavior (CWB)~\cite{sackett2002structure}. 
Currently, the most relevant scope of insider threat is maintained by the CERT division of Software Engineering Institute at Carnegie Mellon University (referred to in this paper as CERT) and is derived from a database of more than 1000 real case studies. 
We refer the reader to a discussion of different perspectives on insider threats in Appendix~\ref{appendix:perspectives}.

\vspace{-0.1cm}
\paragraph{\textbf{Ethical Question of Whistle-Blowers}}
As we have already mentioned, whistle-blowers  usually act due to their loyalty to the society and to the country, while at the same time they break official secret act or prior consent about the confidentiality of information.
We do not take a side of this question, instead, we present this kind of insider threat as part of the state-of-the-art taxonomies that put it under malicious insider threat. 
We utilize such an assignment in our proposed structural taxonomy of insider threat incidents purely for incident classification purposes.

\subsection{\textbf{Definitions}}
Most of the following definitions distinguish between insiders and insider threat. 
Insider definitions usually refer to static descriptions of individuals, using terms such as access, knowledge, trust, or security policy, while insider threat definitions refer to corresponding actions such as misusing access or knowledge which insiders have or violation of a security policy.
Note that the distinction between malicious and unintentional insider types only makes sense for definitions of insider threat, as they require that some action be performed. 
Most existing definitions of insider threat implicitly assume a malicious intent of this threat~\cite{2005-Theoharidou, 2001-Schultz,www-2010-Einwechter}. 
However, other existing definitions do not distinguish between a malicious and unintentional insider and may cover both~\cite{2005-Bishop-2,2011-Hunker,2010-Pfleeger,2008-Predd,2010-Greitzer}. 
Only one of the studies explicitly define unintentional insider threat~\cite{2016-Collins-CERTv5}.

\subsubsection{\textbf{Term Insider}}

Pfleeger et al.~\citeyear{2010-Pfleeger} define an insider as ``a person with legitimate access to an organization's computers and networks.''
The report from RAND Corp.~\cite{2004-Brackney} defines an insider as ``an already trusted person with access to sensitive information and information systems'' (IS).
Aimed at database security, Garfinkel et al.~\citeyear{2002-Garfinkel} consider an insider ``as a subject of the database who has personal knowledge of information in the confidential field.''
Chinchani et al.~\citeyear{2010-Chinchani} attribute the term insider to ``legitimate users who abuse their privileges, and given their familiarity and proximity to the computational environment, can easily cause significant damage or losses.''
According to Althebyan and Panda~\citeyear{2007-Althebyan}, an insider is ``an individual who has the knowledge of the organization's IS structure for which he/she has authorized access and who knows the underlying network topologies of the organization's IS.''
Sinclair and Smith~\citeyear{2008-Sinclair} define an insider as ``any person who has some legitimate privileged access to internal digital resources, i.e., anyone who is allowed to see or change the organization's computer settings, data, or programs in a way that arbitrary members of the public may not.''
A trust-based definition of an insider by the Dagstuhl seminar on Countering Insider Threat~\cite{2008-Probst} concluded that an insider ``is a person that has been legitimately empowered with the right to access, represent, or decide about one or more assets of the organization's structure.''
Greitzer and Frincke~\citeyear{2010-Greitzer} define the insider as ``an individual currently or at one time authorized to access an organization's IS, data, or network.''
Bishop~\citeyear{2005-Bishop-2} determines the insider definition with regard to security policy that contains specified rule set.
He defines an insider as ``a trusted entity that is given the power to violate one or more rules in a given security policy.''
A non-binary approach, indicating ``\textit{degrees of insiderness}'' with access control rules used to develop these degrees, was proposed by Bishop et al.~\citeyear{2009-Bishop}, who defined someone as ``an insider with respect to access to some well-defined data or resource.''
According to Predd et al.~\citeyear{2008-Predd}, an insider ``is someone with legitimate access to an organization's computers and networks.''
Note that the authors deliberately do not define the meaning of word \textit{legitimate}, hence they do not separate insiders from outsiders; instead, they state that ``both legitimate access and the system's perimeter are a function not only of system-specific characteristics but also of a given organization's policies and values.''
Therefore, an insider might also be represented by an external entity such as contractor, ex-employee, business partner, etc.~\cite{2008-Predd}.

\subsubsection{\textbf{Term Insider Threat}}
Pfleeger et al.~\citeyear{2010-Pfleeger} defines insider threat as ``an insider's action that puts at risk an organization's data, processes, or resources in a disruptive or unwelcome way."
According to Greitzer~\citeyear{2010-Greitzer}, the insider threat refers to ``harmful acts that trusted insiders might carry out; for example, something that causes harm to an organization,
or an unauthorized act that benefits the individual.''
According to  Theoharidou et al.~\citeyear{2005-Theoharidou}, insider threat refers to ``threats originating from people that have been given access rights to an IS and misuse their privileges, thus violating the IS security policy of the organization.''
Insider threat is defined by Hunker and Probst~\citeyear{2011-Hunker} as ``an individual with privileges who misuses them or whose access results in misuse.''
Schultz and Shumway~\citeyear{2001-Schultz} define an insider attack as ``the intentional misuse of computer systems by users who are authorized to access those systems and networks.''
Bishop~\citeyear{2005-Bishop-2} defines insider threat as an event occurring when ``a trusted entity abuses the given power to violate one or more rules in a given security policy.''
According to Predd et al.~\citeyear{2008-Predd}, insider threat is ``an insider's action that puts an organization or its resources at risk.''

\subsubsection{\textbf{Term Unintentional Insider Threat}}
According to Collins et al.~\citeyear{2016-Collins-CERTv5}, an unintentional insider threat is defined as ``a current or former employee, contractor, or other business partner'' who: 1) ``has or had authorized access to an organization's network, system, or data,'' and 2) ``had no malicious intent associated with his/her action (or inaction) that caused harm or substantially increased the probability of future serious harm to the confidentiality, integrity, or availability of the organization's information or IS.''
Liu et al.~\citeyear{2009-Liu-2} consider inadvertent insider threat to be defined as ``inattentive, complacent, or untrained people who require authorization and access to an IS in order to perform their jobs.''
Raskin et al.~\citeyear{2010-Raskin} introduce the notion of ``\textit{unintended inference}, which addresses what is not explicitly said in the public text made by an insider,'' and thus he/she may inadvertently reveal some private information.

\subsection{\textbf{Unintentional Insider Threat Taxonomies}}

\paragraph{\textbf{Taxonomy by CERT}}
Greitzer et al.~\citeyear{2014-Greitzer} define four types of unintentional insider threat (derived from the Privacy Rights Clearinghouse):
\textbf{``malicious code''} -- sensitive information social engineered (e.g., planted USB drive,  phishing attack) in combination with malware or spyware;
\textbf{``disclosure''} -- ``sensitive information posted publicly on a Web or sent to unauthorized recipients via fax, mail, or email;''
\textbf{``improper/accidental disposal of physical records''} -- ``lost, discarded, or stolen non\--elec\-tronic records, such as paper documents;'' and
\textbf{``portable equipment no longer in possession''} -- ``lost, discarded, or stolen data storage device, such as a laptop, PDA, smart phone, portable memory device, CD, hard drive, or data tape.''

\paragraph{\textbf{Typologies of Insiders.}}
Considering motivation, Wall~\citeyear{2013-Wall} states that unintentional insider threats are divided according to previous research into \textbf{well-meaning} and \textbf{negligent}.
Then, the author presents the typology of four risk types of employees who can cause data spillage: \textbf{underminers}, (those that make their lives easier by not respecting security policies), \textbf{over-ambitious}, (those that purposely bypass security measures in order to be more effective), \textbf{socially engineered}, and \textbf{data-leakers}.
The author outlines various ways in which data can be leaked, such as accidentally (losing USB sticks); for user's convenience (copy data to home PCs); among others (data present on the hard drives of discarded PCs, sharing with third parties, leakage through public~post,~etc.).

\paragraph{\textbf{Taxonomy of Human Errors}}
Reason~\citeyear{reason1990human} defines human error as ``the failure to achieve the intended outcome in a planned sequence of mental or physical activities when failure is not due to chance.'' 
The author proposes a generic error modeling system (GEMS). 
GEMS divides errors into: \textbf{``slips''}, which represent ``the incorrect execution of a correct action sequence'' (execution failures), and  \textbf{``mistakes''}, which represent ``the correct execution of an incorrect action sequence'' (planning failures).

\subsection{Malicious Insider Threat Taxonomies}

\paragraph{\textbf{Inside Misusers}}\label{TAX-Anderson}
One of the earliest classifications of internal misuse of computer systems was proposed by Anderson~\citeyear{1980-Anderson} who distinguishes among three types of illicit inside users, ordered by the ascending difficulty of their detection from audit trails:
\textbf{masqueraders} -- who can be either external attackers that bypassed security controls and penetrated into a computer system, or internal users who intend to exploit another user's credentials in order to  perform some malicious action;
\textbf{misfeasors} -- 	users who do not masquerade, but instead abuse their own privileges in order to misuse the system; and	
\textbf{clandestine} users -- who represent superusers with the capability of staying under the radar of security controls, which they manage and thoroughly know, and making them most difficult to detect.
Note that internal misuse of a computer system may also include some activities that fall into the category of counterproductive workplace behavior, which are not covered by any of the insider threat definitions.
Therefore, we consider insiders misusing computer systems as the superset of malicious insider threat.

\paragraph{\textbf{Insider Attack Types}}
Bellovin~\citeyear{2008-Bellovin}  distinguishes among three types of insider attacks:
\textbf{the misuse of access} that is, according to the author, the hardest type of incident to detect, because an insider uses a legitimate access for the improper purpose;
\textbf{defense bypass} where insiders have already passed some lines of defense (e.g., firewalls) and can also bypass others; and
\textbf{access control failure} that represents a technical problem, where either vulnerabilities are present in an access control element or such an element is misconfigured.

\paragraph{\textbf{Insider Threat with Various Types of Knowledge}}
Salem et al.~\citeyear{2008-Salem} divide malicious insider threats into two groups: \textit{traitors} and \textit{masqueraders}.
These two classes can be distinguished based upon the amount of knowledge they have.
\textbf{Traitors} have full knowledge about the systems they work with on a daily basis, as well as the actual security policies.
Traitors usually act on their own behalf, and therefore use their own credentials for malicious actions.
On the other hand, \textbf{masqueraders} may have far less knowledge than traitors.
They are attackers who steal the credentials of another legitimate user, and then use the stolen credentials for executing a malicious act on behalf of another user.
Another example is obtaining an access to the account of a victim by exploiting some system vulnerability.
Note that these two classes are not necessarily disjoint; traitors can first use their legitimate access for obtaining another user's access and then perform damage using that access.

\paragraph{\textbf{Categorization of Insider Threat by Network Monitoring.}}
Myers et al.~\citeyear{2009-Myers} divide malicious insider threat into two categories from the perspective of network event monitoring:
\textbf{``the unauthorized use of privileges,''} representing insiders who access data that they are not authorized to access (i.e., data that is not related to the project or role of an insider), or misuse their authorized access in an inappropriate way (e.g.,
sharing classified data with an unauthorized person); and
\textbf{``automated insiders''} that represent either bots performing reconnaissance of the internal network, or probes identifying vulnerabilities and misconfigurations in the internal systems.

\paragraph{\textbf{Categorization of Insider Threat from the Cloud Computing Perspective.}}
Claycomb and Nicoll~\citeyear{2012-Claycomb} categorize insider threat with regard to cloud computing into three categories:
\textbf{the rogue cloud provider administrator} who can access a potential victim's data or other leased resources that can be leaked or misused for fraud, IP theft, or sabotage of the cloud provider's reputation;
\textbf{insiders who exploit cloud vulnerabilities for unauthorized purposes}, e.g., exploitation of replication lag among replicas for fraud; and
\textbf{insiders who exploit the cloud services} in order to conduct nefarious activity against the company (e.g., cracking password files, DDoS attacks against an employer, exfiltration of data when leaving the company).
In contrast to previously mentioned categories, targeted systems or data are not necessarily directly related to cloud services.
Another categorization of insider threat from cloud computing perspective is presented by Kandias~et al.~\citeyear{2011-Kandias} who distinguish between a) an \textbf{``insider employed by the cloud provider''}, and b) an \textbf{``insider employed by an organization''} that outsources IT to the cloud.

\paragraph{\textbf{Classification of Insiders.}}
Cole and Ring~\citeyear{2006-Cole} classify insiders into four classes:
1) A class of \textbf{pure insiders}, which contains regular employees that only have privileges necessary to perform their jobs (door keys, access cards, access to a specified list of network services, etc.).
A special case is \textit{``an elevated pure insider''} who has privileged access.
2) A class of \textbf{inside associates} (a.k.a. external insiders~\cite{2010-Franqueira}) consists of third party personnel, such as contractors and suppliers, as well as internal employees with limited authorized access to various compartments inside an organization, for example, security guards, servicemen, and cleaners.  	
The limited authorized access of inside associates is usually represented by physical access to the department/facility, as opposed to access to the IT infrastructure of an organization.
Inside associates, such as cleaners, have physical access to the workspace of other employees, and therefore they may find privacy-sensitive information on employees' desks or in trash bins. 
Moreover, they may plant key-loggers for the purpose of sniffing credentials or other sensitive information entered by employees using keyboards.  
In comparison to the previous classes, \textbf{3) inside and 4) outside affiliate} classes represents people that do not have any justified and legitimate reason to enter the building of an organization.
An inside affiliate class involves family members, friends, or clients of an employee. 
They have no access to the organization's facility but can steal and misuse the access cards/credentials of an employee to obtain such access and further perpetrate malicious acts.
The class of outside affiliates includes untrusted people external to an organization, who can obtain internal access to the organization's network by, for example, using unprotected WiFi, bypassing weakly protected WiFi, or social engineering the credentials of authorized employees (e.g., phishing attacks).

\paragraph{\textbf{High-End and Low-End Profiles of Insiders.}}
Cole and Ring~\citeyear{2006-Cole} suggest profiling insiders into two categories, taking into account the actions, appearances, and instincts of co-workers:
\textbf{1)~low-end insiders} -- this profile includes the characteristics of the insiders that have been caught, accused, and potentially convicted, while also keeping in mind that the majority of high-end insiders are smarter and have not been caught.
The basic characteristics of low-end insiders are as follows:
they have no or minimal technical education/knowledge;
they have already worked in a variety of positions;
the attacks their perpetrate involve theft of IP;
their primary motivation is personal gain;
they are unaware of potential negative consequences of their acts;
their wrong doing causes an increased attention and suspicion of their colleagues;
they are influenced by their emotional state  (e.g., anger or grievance).
\textbf{2) high-end insiders} -- this profile of insider involves people who usually view their malicious mission as their career decision, in contrast to low-end insiders whose actions are usually short-term.  
Therefore, the authors consider high-end insiders as more dangerous than low-end insiders.
The high-end insider profile has, according to the authors, several common characteristics:
they are diligent employees,
they want to achieve high career positions quickly, while evincing very good leadership skills, job dedication, and trustworthiness.

\paragraph{\textbf{Levels of Insider Threats}}
Based on the various potential consequences and harm to an organization, Cole and Ring~\citeyear{2006-Cole} distinguish among three levels of insider threats:
\textbf{Level 1: Self-motivated insiders} are not motivated by any third party; instead they decide to act on their own due to some personal reason, such  as revenge, need for correction of organization's actions (e.g., whistle-blowers), or complacence.
\textbf{Level 2: Recruited insiders} are those who do not independently decide to act maliciously but are convinced and hired by third party.
Usually, these insiders are successfully recruited due to  financial problems or other personal weakness that can be exploited by a third party.	
Because this type of recruited insider can be risky for both the insider and the recruiter, the preferred way for malicious third parties is to plant their own insider (e.g., spy) into a target victim's company.
\textbf{Level 3: Planted insiders} 
are placed by malicious organizations that find someone suitable for an insider job, train them, get them hired at the target victim's company, allow them time to earn trust within the company, and then start exploiting the insider for data exfiltration/espionage.

\paragraph{\textbf{Types of Motivations}}
According to Cole and Ring~\citeyear{2006-Cole}, there are plenty of factors that may contribute to the motivation of an insider who ``turns to the dark side,'' however there are three main motivations that appear repeatedly:
\textbf{financial} -- when organizations are recruiting people to perform inside attacks, they may target people with financial difficulties or those who want to earn some extra money, which may also be the reason for a self-motivated insider;
\textbf{political} -- some employees have strong political opinions, and if their employers have substantially different opinions and actions, then these employees are motivated to cause harm when opportunity comes or to commence collaboration with malicious entities outside the organization; and
\textbf{personal} -- this type of motivation can occur in one of two ways: 1) the recruiter digs into the victim's past and tries to find out the victim's deepest, darkest secrets and use them to threaten the victim (e.g., blackmail), or 2) if the victim has no secrets accessible to the recruiter, then the recruiter tries to orchestrate a trap scenario that will create a new secret (e.g., using an attractive human bait).

\paragraph{\textbf{Insider Profiling by CERT}}\label{TAX:profiling-by-CERT}
Cappelli et al.~\citeyear{2012-Cappelli} propose profiling malicious insider threats into three types:
\textbf{1) IT sabotage} in which ``an insider uses IT to direct specific harm at an organization or an individual.''
Such insiders are usually disgruntled employees with technical background who have administrative privileges.
An example of this category is the installation of a logic bomb that is activated after an employee's termination.
\textbf{2) Theft of Intellectual Property}.
Common cases involve espionage, and they are usually committed by technical staff (e.g., engineers and developers), as well as non-technical staff (e.g., clerks and salesmen).	
Perpetrators may steal information that they daily access, and take it with them when they leave the organization (e.g., using the IP for their own business, taking it to a new employer, or passing it to another organization).
\textbf{3) Fraud} in which ``an insider uses IT for the unauthorized modification, addition, or deletion of an organization's data for personal gain or theft.''	
Insider fraud is usually perpetrated by low-level staff with non-technical backgrounds, such as human resources or help desk staff.	
The motivation for this is often greed or financial difficulties, and this type of crime is typically long-term.
Recruitment of such fraudsters by external entities is also very common. 
The case studies that do not fall under these three profiles are denoted as miscellaneous~\cite{2016-Collins-CERTv5}.

\subsection{Combined Taxonomies of Insider Threat}

\paragraph{\textbf{Categorization of Insiders from Hayden.}}
Hayden~\citeyear{1999-Hayden} categorizes insider threat into four groups: ``\textit{traitor, zealot, browser,} and \textit{well intentioned,}'' where the former two are related to malicious insider threat and the latter two to unintentional insider threat.
The \textbf{traitor} group consists of people ``who have malevolent intent to damage, destroy, or sell out their organization.''
The \textbf{zealot} group refers to people ``who believe strongly in the correctness of one's position or feel the organization is not on the right side of a certain issue.''
In this case, the insider may try to \textit{``correct the organization,"} for example, by disclosure or removal of confidential data, or providing access to unauthorized parties.
The \textbf{browser} category refers to people ``who are overly curious in nature, often violating the need-to-know principle.''
The \textbf{well-intentioned} category of insiders involve people who ``commit violations through ignorance, e.g., disabling anti-virus protection, using unapproved thumb drives.''

\paragraph{\textbf{Policy-Based Taxonomy of Insider Threat}}
A policy-based taxonomy of insider threat can be derived from the definitions and corresponding model of insider threat proposed by Bishop and Gates~\citeyear{2008-Bishop}.
The authors  expanded their former definition of insider threat~\cite{2005-Bishop-2} by arguing ``that a security policy is inherently represented by the access control rules employed by an organization.''
Then, an insider threat can be categorized considering two primitive actions:
\textbf{``violation of a security policy using legitimate access,''} involving actions that do not respect security policy in force, for example, confidential data is leaked to an unauthorized person; and
\textbf{``violation of an access control policy by obtaining unauthorized access''} -- in this case, insiders misuse their legitimate access in order to extend the scope of their actual privileges, while breaking the security policy in force as well as the access control mechanism, e.g., gaining superuser access by exploiting some system vulnerability.
Note that authors do not explicitly mention an intent of the insider threat, therefore it may cover malicious as well as unintentional threats.
With this in mind, an example of gaining superuser access can be made for the purpose of stealing confidential data or for the purpose of making the work more efficient, while disobeying policies.

\paragraph{\textbf{Categorization of Inside Misusers by Physical Presence}}
Considering physical presence, Neumann~\citeyear{2010-Neumann} classifies insiders into two categories: logical and physical.
A \textbf{logical insider} executes his/her actions physically outside of an organization's workspace, while a \textbf{physical insider} acts inside of the physical boundaries of the organization's workspace, including external trusted networks.
Note that this taxonomy does not specify the intent of the insiders, and therefore can be applied to both unintentional and malicious insider threats.

\paragraph{\textbf{Human-Centric Taxonomy of Inside Misusers}}
Magklaras and Furnell~\citeyear{2002-Magklaras} propose a human-centric taxonomy of inside misusers considering three dimensions: \textit{insider's role in the system}, \textit{reason for misuse}, and \textit{consequences on the system}.
1) \textbf{System role} --
the first dimension classifies people by ``the type and level of system knowledge they possess.''
This dimension consists of:
\textit{system masters} who have full control over most of the IS resources (e.g., system administrators);
\textit{advanced users} who, although do not have privileged access, acquired a large amount of knowledge about the systems and networks of the organization, and also are often capable of revealing system vulnerabilities (e.g., programmers, database administrators); and
\textit{application users} who use certain standard	applications, like Internet browsers, office suites, and email clients, but usually do not have extra privileges to access resources other than those required by their applications.
2) \textbf{Reason for misuse} -- 
this dimension describes attributes of insider threat incidents.
Considering this dimension, the authors categorize insiders into two groups:
\textit{intentional} misfeasors who act for various reasons (e.g., deliberate ignorance, revenge) and 
\textit{accidental} misusers who can also be classified by the actual reason that negatively influenced the behavior of a legitimate user (e.g., lack of training, excessive workload, personal problems).
3) \textbf{System consequences} -- 
this dimension distinguishes among the various ways a misuse act occur, which is manifest by certain traces in the IT infrastructure at the system level.
The authors describe three levels that are attributed to these consequences:
\textit{OS} -- 
modifications to the structure of a file system, the installation of unauthorized software, etc.;
\textit{network} -- network packets may contain unauthorized content, data exfiltration of confidential data may be perpetrated through email or file sharing services, etc.; and
\textit{hardware} -- vandalism or removal of hardware components, installation of key-loggers, modifications of default configurations to critical hardware components (e.g., for the purpose of sabotage or IP theft).

\paragraph{\textbf{Detection-Oriented Taxonomy of Inside Attacks}}
Similar to the system consequences dimension proposed by Magklaras and Furnell~\citeyear{2002-Magklaras}, Phyo and Furnell~\citeyear{2004-Phyo-1} propose a taxonomy of inside attacks, which distinguishes among four monitoring levels of a target system at which an attack may be detected:
1) \textbf{network}, 2) \textbf{operating system}, 3) \textbf{application}, and 4) \textbf{data}.
This taxonomy is based on the assumption that one inside attack may be manifest at specific levels of the system, while traces of another inside attack may be present at different levels
(e.g., fraud breaching integrity of data may be obvious at application and data levels, while data exfiltration may be manifest at network and OS levels).
Note that this taxonomy does not discern the intent of the insiders, hence it can be inherently applied to both malicious and unintentional insider threats.

\paragraph{\textbf{Subtypes of Perpetrators}}
The typology of insider perpetrator subtypes is presented by Shaw and Fischer~\citeyear{2005-Shaw} and includes two subtypes representing unintentional insider threat and six subtypes representing malicious insider threat.
This typology includes:
\textbf{explorers} who are inquisitive people (largely with benign intent) that violate policies while they explore the system and its components;
\textbf{samaritans} who are individuals that do not follow approved procedures for fixing some issues, but instead ``hack into a system'' in a more efficient or faster way;
\textbf{hackers} who are individuals that, despite having records of previous hacking activities, are employed and continue in these activities.
They may also install logic bombs that serve as employment insurance in cases in which their malicious activities are discovered;
\textbf{machiavellians} who are individuals that perpetrate sabotage, industrial espionage, intellectual property theft, or other types of activities that potentially lead to their job promotion (e.g., intentionally introduce critical bugs that are discovered by these \textit{triumphant} insiders);	
\textbf{proprietors} who are individuals with a strong feeling of possession of the systems they are managing and are ready to defend ``the control and power over this territory'' -- in many cases these insiders prefer to destroy some critical components of the system or the system itself instead of giving up;
\textbf{avengers} who are disgruntled individuals that act impulsively due to perceived injustice against them;
\textbf{career thieves} who are individuals that start their job exclusively with the intention of perpetrating malicious acts that they can benefit from, such as financial fraud or the theft of intellectual property;
and \textbf{moles} who are individuals that start their job exclusively with the intention of stealing intellectual property for the benefit of a foreign country, organization, or competing company.

\paragraph{\textbf{Insider Taxonomy from Sinclair}}
Sinclair and Smith~\citeyear{2008-Sinclair} divide the insider threat into three classes: insiders with an intention to commit malicious action; insiders acting for their personal profit; and insiders who accidentally or unwittingly act in a harmful way.

\paragraph{\textbf{Categorization of Insider Threats using Trust}}
Probst and Hunker~\citeyear{2010-Probst-2} consider trust and categorize insider threats into two categories:
\textbf{those not representing a violation of trust}, which consist of \textit{accidental insider} threats and \textit{disobeying a security policy} in force; and
\textbf{insider threats representing a violation of trust}, which include \textit{simple insider} and \textit{high profile insider} threats (similar to the low-end and high-end insiders in~\cite{2006-Cole}).

\subsection{\textbf{Structural Taxonomy of Insider Incidents by 5W1H}}
Taking into account the definitions and taxonomies introduced above, in this section we present our structural taxonomy of insider threat. 
In order to provide a unified view regarding the existing taxonomies, we utilized the \textit{who, what, where, when, why}, and \textit{how} (5W1H) methodology.
\begin{figure*}
	\vspace{-0.1cm}
	\centering
	\includegraphics[width=1.00\textheight,angle=-90,origin=c]{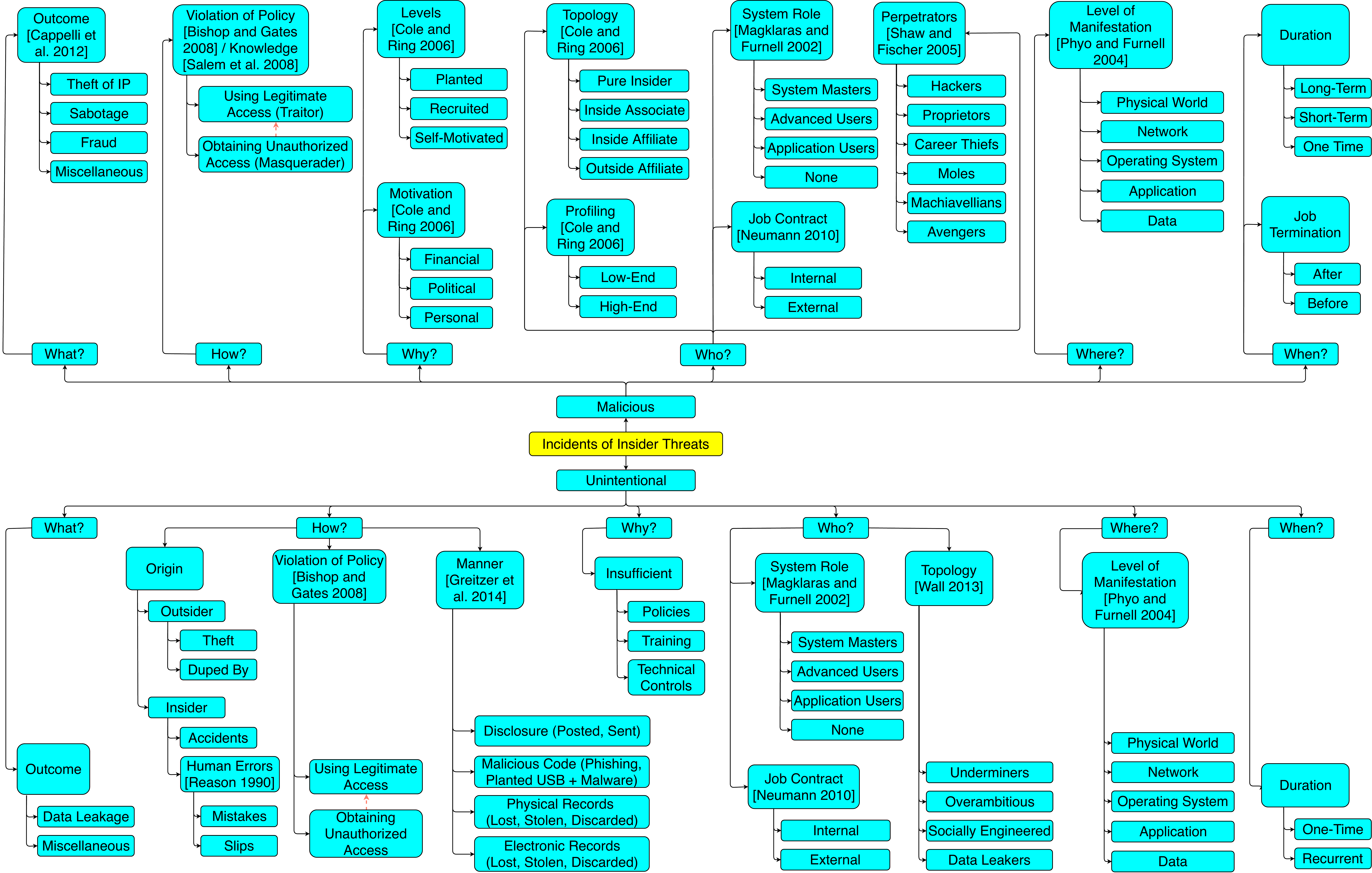}
	\vspace{-0.2cm}
	\caption{Structural taxonomy of insider threat (based on previous research and 5W1H questions)\label{fig:unification-of-taxonomy}}
\end{figure*}
The 5W1H are elementary questions addressing the information gathering problem, which were originally used to report a news story, but have had other applications as well (e.g., building domain ontologies~\cite{yang20115w1h}). 
Since the insider incident investigation problem can also be viewed an instance of information gathering, we selected this approach for a structural taxonomy of this type of incident.

First, we selected several sample taxonomies of insider threat answering 5W1H questions with regards to new information brought.
In the selection, we targeted taxonomies that are mostly orthogonal, while we skipped ones that are specific to a particular domain (e.g., \cite{2012-Claycomb}) or that overlap with other more broad taxonomies (e.g., \cite{2008-Sinclair}).
Then, according to the remaining relevant answers to the 5W1H questions, we supplemented our selection with additional further subcategorization that can enhance the description of a particular insider threat incident in a systematic manner.
The proposed schema of structural taxonomy is depicted in Figure~\ref{fig:unification-of-taxonomy} (the left side of the figure is related to malicious insider threat, and the right side is related to unintentional insider threat).
In addition to assigning some of described taxonomies to 5W1H questions, we subcategorized malicious insider threat incidents by adding two ways of addressing the question of \textit{when}; the first one describes duration, and the second differentiates between whether the incident was performed before or after an insider left the place of his/her employment (i.e., job termination).
For the sake of simplicity, this inherently assumes that every malicious insider is fired by an organization or decides to leave a place of employment on his/her own. 
The next two additional subcategorizations were added to the unintentional insider threat incidents; the first one relates to the question of \textit{why}, and the second one relates to the question of \textit{when}.
Note that after distinguishing between malicious and unintentional insider threat, some of the taxonomies can be based on an abstract level reduced into one (e.g.,~\cite{2008-Salem} and ``malicious'' part of taxonomy derived from~\cite{2008-Bishop}). 
In the case of subcategorization by a violation of policy, we depicted the requirement of obtaining unauthorized access based on the use of legitimate access with a red dashed arrow.
By using the proposed structural taxonomy, all important information about an insider threat incident is kept in an easily maintained and clear format, which can be extended or modified in a straightforward manner in order to integrate future case studies.
Two examples that demonstrate the application of our proposed structural taxonomy are presented in Appendix~\ref{appendix:structural-categorization-examples}. 

\section{Proposed Categorization}\label{Sec:Proposed-Categorization}
After surveying definitions and taxonomies related to insider threat, we now change our focus to the second main contribution of this work -- our proposed novel categorization of existing insider threat research.
In this section we present an overview of the four main categories, and we subsequently expand on each of the categories in dedicated sections below.

\subsection{\textbf{Workflow of Research Contributions}}
For the categorization and review of all of the studies included in our survey (and respectively their contributions), we applied the grounded theory approach for rigorous literature review ~\cite{wolfswinkel2013using}, which consists of five stages.
In the initial stage, several inputs are specified: the criteria for inclusion/exclusion, field of research, appropriate data sources, and search terms.
Searching is performed in the second stage, followed by the stage of selecting, which includes filtering out doubles, refining samples based on titles and abstracts (later based on full texts), and inclusion of forward and backward references.
The fourth stage is aimed at analysis, and it applies the key principles of the grounded theory approach: identification of high level categories from found concepts (open coding); identification of subcategories among the high level categories (axial coding); and refinement and integration of existing categories and their subcategories (selective coding).
The last stage deals with presenting the findings and insights in the area.

In addition to the application of grounded theory, our intention was to depict the workflow among particular categories of the contributions that would follow the direction from incidents to solutions or vice versa.
In this process, we identified step-by-step \textit{defense solutions} for the detection, assessment, prevention, and mitigation of insider threats; these solutions were further subdivided to discern \textit{simulation} approaches as an independent category.
Afterward, we identified research contributions that either analyze and model the behavior of an insider threat, or study its psychosocial precursors; then we created a category of such contributions and called it \textit{analysis of incidents}.
Finally, we established a separate category for insider threat \textit{incidents and datasets}, which involves collections of miscellaneous datasets of cyber observable data and real-world case studies.
\begin{figure}[t]
	\centering
	\vspace{-0.4cm}
	\includegraphics[scale=0.65]{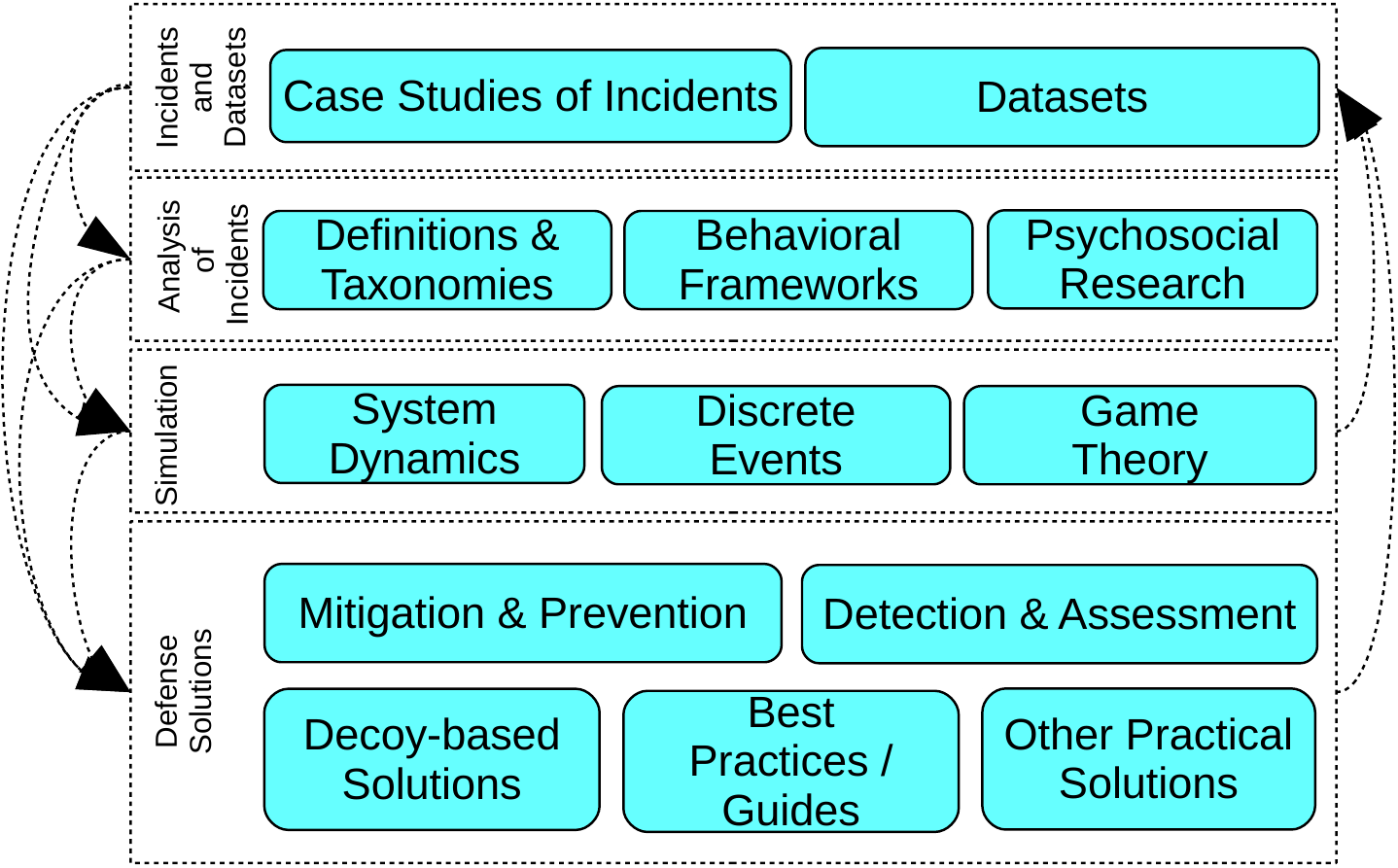}
	\vspace{-0.2cm}
	\caption{Workflow of research contributions\label{research-workflow}}
	\vspace{-0.5cm}
\end{figure}
Thus, the proposed workflow of insider threat contributions consists of four main categories:
\begin{compactitem}
	\item The \textbf{Incidents and Datasets} category contains reference \textit{datasets} applicable for the evaluation of insider threat detection approaches, as well as collections describing real-world insider \textit{incident case studies}, which can be utilized for the analysis and modeling of insider threat or the design of defense solutions aimed at specific types of incidents.
	More information about this category can be found in Section~\ref{Sec:Incidents}.
	\item The \textbf{Analysis of Incidents} category aims at generalization and modeling of all related aspects and behaviors of insider threat incidents and includes \textit{definitions and taxonomies} (previously described in Section~\ref{sec:Taxonomies}), \textit{behavioral frameworks} modeling the insider threat lifecycle, observed indicators and critical pathways from the security perspective, as well as contributions of studies from \textit{psychological and social} areas that also involve criminology theories.
	This category has significant importance in understanding the behavior of the malicious insider and his/her trails and observations, which are indications that malicious activities have begun.
	This information should be taken into account when designing defense solutions.
	Detailed information about this category can be found in Section~\ref{Sec:Analysis-of-Incidents}.
	\item The \textbf{Simulations} category includes research contributions that perform experiments with programmed models of simulated insider environments, either for the purpose of investigating the impact of various simulation settings on the execution of a simulation, or for the purpose of synthetic dataset generation that may be used for testing defense solutions.
	The simulation category contains three subcategories: \textit{system dynamics}, \textit{discrete event simulation}, and \textit{game theory} works.
	Further information regarding this category is available in Section~\ref{Sec:Simulations}.
	\item The \textbf{Defense Solutions} category is the largest category and includes contributions that propose a solution for the insider threat \textit{detection} \textit{assessment}, \textit{prevention, or mitigation}; we include a special subcategory for \textit{decoy-based solutions} (such as honeypots and honeytokens) and procedural defense solutions in the form of \textit{best practices and guides}.
	Moreover, for the sake of comprehensiveness, this category also contains various \textit{other practical defense solutions} involving several commercial tools.
	In sum, this category provides knowledge about the spectrum of the defense options and reveals trends and ideas in the development of defense solutions. Section~\ref{Sec:Defense} provides a detailed description of this category.
	
\end{compactitem}
The workflow-based relations of insider threat categories, along with their subcategories, are depicted in Figure~\ref{research-workflow}. 
In this figure, the top-down direction represents the direction from incidents to solutions (arrows on the left side of the diagram), and the bottom-up direction represents the workflow from solutions to incidents (arrows on the right side of the diagram).
The direction from incidents to solutions (top-down) represents a goal-oriented workflow with the ultimate goal of the design and development of defense solutions.
In this paper, we followed this order of the workflow in order to model and present the categorization.
More specifically: all categories are dependent on the incidents and datasets category; 
the analysis of incidents can be used for the purpose of the simulation and design of defense solutions;
defense solutions can also utilize the results of the experiments from simulation, and thereby improve the settings of the detection solutions. 

On the other hand, in some cases the workflow may go in the opposite direction -- from the bottom to top.
In this case,  simulations  may contribute to the generation of synthetic datasets.
Best practices related to detection may extend the knowledge about real case studies (as more incidents can be detected), while detection can produce real datasets as a testbed for other defense solutions.
A more detailed categorization of the research contributions is presented in Appendix~\ref{appendix:overal-categorization}.

\section{Incidents and Datasets}\label{Sec:Incidents}
\begin{figure}[t]
	\centering
	\vspace{-0.4cm}
	\includegraphics[scale=0.45]{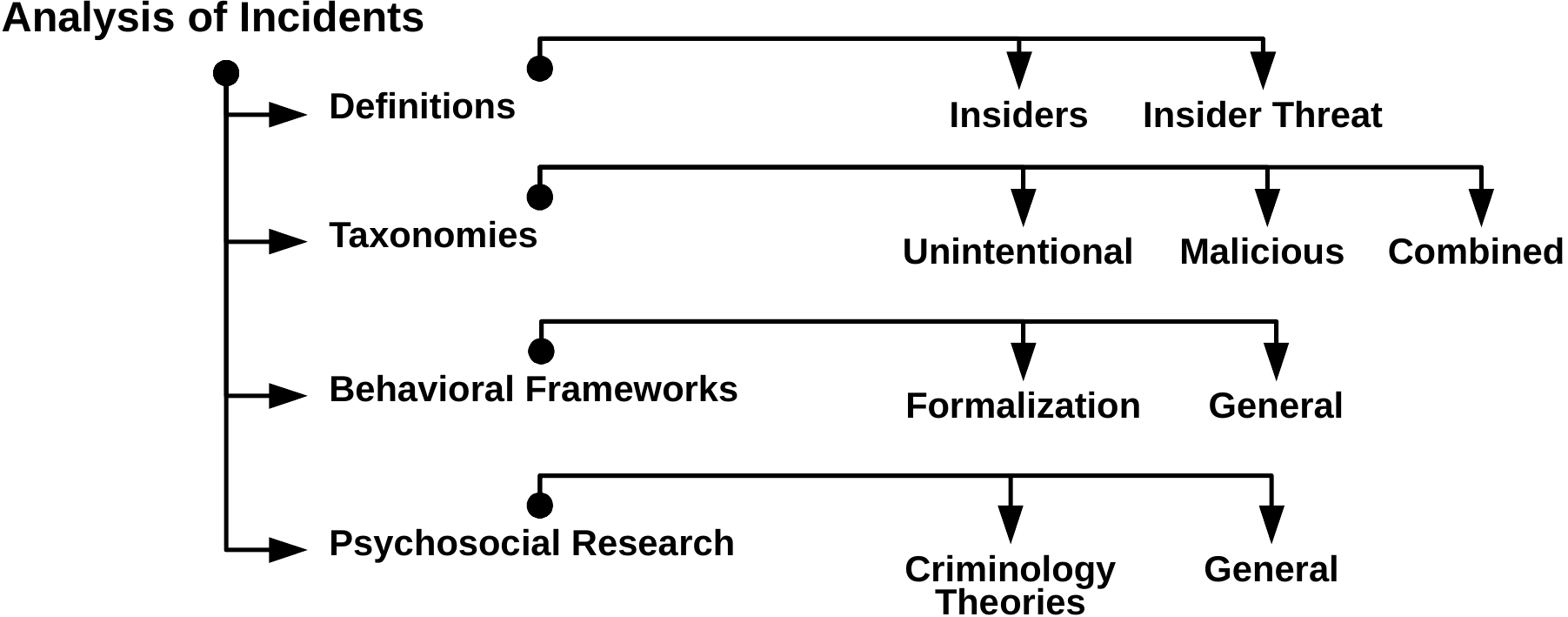}
	\vspace{-0.25cm}
	\caption{Detailed categorization of the \textit{Incidents and Datasets} category\label{fig:categorization-incidents}}
	\vspace{-0.3cm}
\end{figure}

A dataset is a key element for designing and evaluating new ideas and solutions in every applied research field.
In this section we present relevant data about case studies of incidents of insider threat, as well as existing datasets gathered either from laboratory experiments or from the real world.
We believe that the following information will help readers acquaint themselves with various case studies and available datasets in this field.
See Figure~\ref{fig:categorization-incidents} for an overview of this category.

\subsection{Case Studies of Incidents}
This subcategory includes several examples of various case studies of insider threat incidents, and we also include two state-of-the-art assignments of such incidents into clusters. 
The following states research that covers various types of insider threat case studies.

The majority of the research aimed at obtaining information about insider incidents was conducted by CERT and the US Secret Service (e.g.,~\cite{2004-Randazzo,2008-Kowalski}).
For example, Randazzo et al.~\citeyear{2004-Randazzo} examined 23 case studies of insider threat occurring in the finance sector between 1996 and 2002; 15 of the incidents involved fraud, four involved IP theft, and four~involved sabotage of the IS/network.
In addition, Kowalski et al.~\citeyear{2008-Kowalski} described 36 insider threat case studies in the government sector that occurred between 1996 and 2002; 21 of the incidents involved various types of fraud, nine involved sabotage, three involved IP theft, and three involved a combination of sabotage and IP theft.
Fischer described several case studies of insider computer abuse of US defense systems in~\citeyear{2003-Fischer}. 
Shaw et al.~\citeyear{1998-Shaw} described several case studies of critical information technology insiders (CITI), including system administrators, programmers, and network professionals.
Sabotage using IT in critical infrastructure sector was also addressed by Keeney et al.~\citeyear{2005-Keeney}.
Shaw and Fischer~\citeyear{2005-Shaw} thoroughly examined 10 case studies of malicious insiders within national critical infrastructure in Washington, DC.
Furthermore, all versions of the CERT guides (e.g.,~\cite{2016-Collins-CERTv5}) provide several descriptions of insider threat case studies; the authors highlight that all insider actions could be prevented by respecting the best practices proposed in the guides. 

Interesting case studies are also described in the works of Magklaras and Furnell~\citeyear{2002-Magklaras}, Jabbour and Menasc{\'e}~\citeyear{2009-Jabbour-2}, Probst et al.~\citeyear{2010-Probst-1}, Bishop et al.~\citeyear{2009-Bishop}, and Predd et al.~\citeyear{2008-Predd}.
All of these authors dealt with either data exfiltration, IP theft, or sabotage, particularly that found in the financial and military sectors.
Moreover, some of them also focused on the unintentional insider threat, e.g., Probst et al.~\citeyear{2010-Probst-1} dealt with phishing attacks, and Predd et al.~\citeyear{2008-Predd} described an episode of unintentional denial of service. 

\paragraph{\textbf{Clustering of Case Studies}}
In their book, Cappelli et al.~\citeyear{2012-Cappelli} described a sample of 51 case studies, which were assigned to five clusters with respect to proposed profiling: 24 involved \textit{sabotage}, three cases contained \textit{fraud with sabotage}, six cases involved \textit{IP theft}, 12 cases involved \textit{fraud} alone, and finally, six cases exhibited a \textit{miscellaneous} character.\footnote{Note that this clustering of case studies corresponds to the taxonomy proposed by these authors -- see Section~\ref{TAX:profiling-by-CERT}.}
In another book, Cole and Ring~\citeyear{2006-Cole} divided the presented case studies into two wide clusters based on the environment of their occurrence: \textit{government} and \textit{corporations}.
The government cluster contains a description of four examples  from \textit{state and local government} and 17 examples from the \textit{federal government}, while in the corporation cluster, the authors present 15 \textit{commercial} examples, eight examples from  the \textit{banking and financial sector}, and two examples of \textit{government subcontractors}.

\subsection{Datasets}\label{Sec:Datasets}
\begin{figure}[t]
	\centering
	\vspace{-0.3cm}
	\includegraphics[scale=0.50]{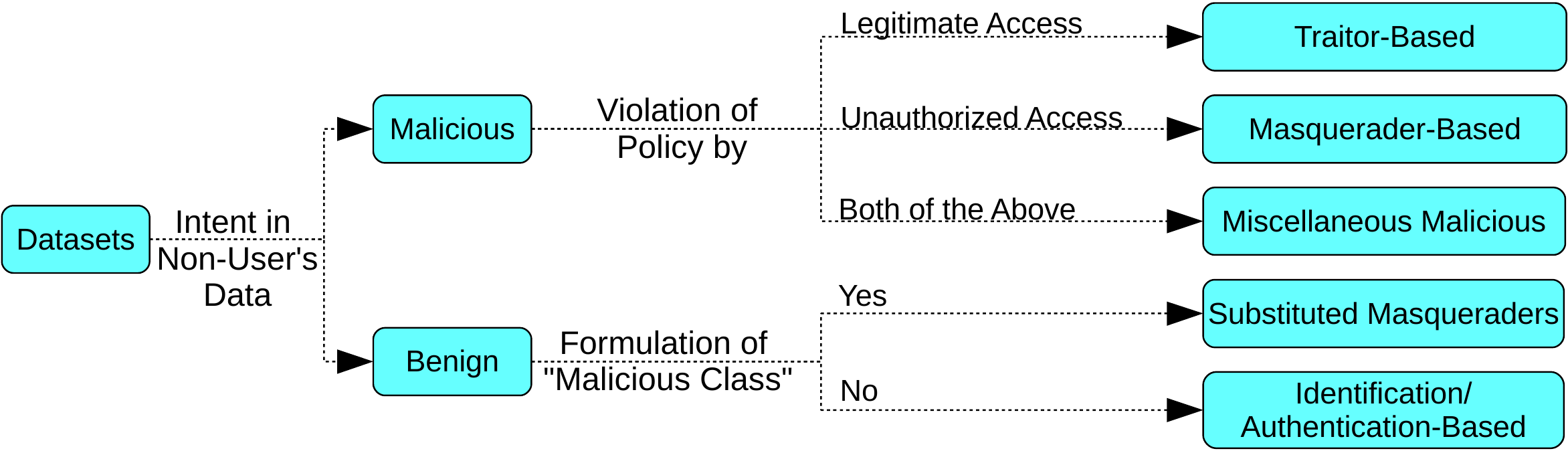}
	\vspace{-0.25cm}
	\caption{Categorization of datasets used in insider threat research \label{Categorization-of-datasets}}
	\vspace{-0.5cm}
\end{figure}

After a review of the insider threat literature, and defense solutions in particular, we divided commonly used datasets into five categories: 1) \textit{masquerader-based}, 2) \textit{traitor-based}, 3) \textit{miscellaneous malicious}, 4) \textit{substituted masqueraders}, and 5) \textit{identification/authentication-based}.
All of these categories are depicted in Figure~\ref{Categorization-of-datasets}, and are also described in our previous work~\cite{2017-twos-mist}.
As can be seen in the figure, these categories can be obtained by applying the following decision steps:
a) by discerning the user's intent in non-user's data (i.e., data not belonging to a user), which yields \textit{malicious} and \textit{benign} branches;
$b_1$) for the malicious intent branch, by the manner in which the policy violation was executed -- either by the use of a legitimate user's access (\textit{traitor-based}), by obtaining unauthorized access (\textit{masquerader-based}), or when both of the cases are included in a dataset separately (\textit{miscellaneous malicious}); and
$b_2$) for the benign intent branch, by discerning whether the malicious class was formulated by authors of a dataset or not, where the \textit{substituted masqueraders} category include dataset with samples containing labels of such an explicitly built ``malicious class'' and the \textit{identification/authentication-based} category does~not -- samples contain only labels of user identification.
Splitting datasets of the benign intent branch into two subcategories enables us to isolate  datasets that specify equal conditions on a detection/classification task, and thereby the evaluation of an approach on these datasets is always reproducible with the same setting. 
In contrast to them, identification/authentication-based datasets enable researchers to select various mixtures of samples for a malicious class and thus potentially simplify the classification task.
Also, note that each of the five dataset classes we proposed can be sub-categorized according to the origin of data into: 1) real-world datasets and 2) laboratory datasets.
However, we are aware only about a single real-world dataset~\cite{dataset-enron}, therefore we use this division criteria only tangentially.

\subsubsection{\textbf{Masquerader-Based Datasets}}
Despite the fact that there has been a significant amount of research conducted dealing with the masquerader detection problem, only a few studies have used datasets specifically designed for this purpose. 
The following examples utilize datasets that contain malicious intent in data entries with malicious labels, and the corresponding malicious scenarios are aimed at violations of policy by obtaining unauthorized access.

\textbf{RUU Dataset}. 
Are You You (RUU) is a masquerader dataset that was introduced by Salem and Stolfo~\citeyear{2009-Salem, 2011-Salem-2}. 
The dataset was collected from the PCs of 34 normal users and consists of host-based events derived from file system access, processes, the Windows registry, dynamic library loadings, and the system GUI. The dataset contains masquerade sessions performed by 14 humans based on the specified task of locating any piece of data that has a direct or indirect financial value; the users were not limited to any particular means or resources.

\textbf{WUIL Dataset}. 
The Windows-Users and Intruder simulations Logs (WUIL) dataset was designed and implemented by Cami\~na et al.~\citeyear{2011-Camina} and contains generic file system interactions regardless of their type (e.g., open, write, read). 
The WUIL dataset contains records from 20 volunteer users (increased to 76 in~\cite{2016-Camina}), who were monitored at different periods of time during their routine daily activities. 
Although, some users produced approximately an hour's worth of logs, others produced logs spanning several weeks. 
The data was collected using an internal tool for file system auditing of Windows machines of various versions (i.e., XP, 7, 8, and 8.1). 
While the legitimate users' data was collected from real users, the masquerade sessions were simulated using batch scripts considering three skill levels of users: \textit{basic}, \textit{intermediate}, and \textit{advanced}. 

\textbf{DARPA 1998 Dataset}. 
The DARPA 1998 Intrusion Detection Evaluation dataset was synthesized by the MIT Lincoln Laboratory based on statistical parameters of a ``government site containing 100's of users on 1000's of different hosts''~\cite{lippmann2000darpa}, and its primary purpose was to evaluate and improve intrusion detection systems.
However, it was also used in research on the insider threat detection problem~\cite{2011-Parveen}.
The DARPA 1998 dataset consists of network traces and system call logs captured on attacked machines, and the attacks performed are divided into four groups: 1) \textit{``denial of service''}, 2) \textit{``remote to user''}, 3) \textit{``user to root''}, and 4) \textit{``surveillance''}.
From the insider threat perspective, the only interesting group of attacks is the \textit{``user to root''} group, which may be viewed as masquerade attacks.
Nevertheless, only system call logs are relevant, as direct consequences of these attacks can only be monitored on the victim hosts.
The DARPA 1998 dataset received a critique~\cite{mchugh2000testing} and currently is considered outdated.

\subsubsection{\textbf{Traitor-Based Datasets}}
In the malicious datasets branch, we identified the masquerader research trend, while research dedicated to traitor detection has been more limited. 
This difference can be explained by the assumption that masquerader detection is simpler and more straightforward than traitor detection, as argued by Salem et al.~\citeyear{2008-Salem} who mentioned that ``a masquerader is likely to perform actions inconsistent with the victim's typical behavior''. 
On the other hand, given that an attacker is a moving target, he/she may imitate a behavior of a victim to some extent, along malicious actions he/she performs.
The following research utilizes datasets that include malicious intent in data considered as malicious, and is aimed at policy violations using legitimate access.

\textbf{Enron Dataset}. 
The Enron dataset~\cite{dataset-enron} consists of a collection of 500,000 real-world emails (from 1998 to 2002) associated with 150 users, mainly senior management of Enron Corporation.
Although, some of the emails were deleted as they contained attachments or confidential information, the dataset contains interesting information that can be used for the analysis of text in emails and social network analysis aimed at the detection of insider threat involving collaborating traitors.

\textbf{APEX 2007}. The APEX '07 dataset was collected, according to Santos et al.~\citeyear{2008-Santos}, by the National Institute of Standards and Technology (NIST) with an intention to simulate the tasks of analysts in the intelligence community.
The APEX '07 dataset consists of actions and research reports of eight benign analysts, while the malicious insider threat was simulated by five analysts whose tasks were based on the tasks of benign analysts in order to make a detection more challenging.

\subsubsection{\textbf{Miscellaneous Malicious Datasets}}

Datasets composed of both malicious insider subtypes (masqueraders and traitors) belong to this category.
Therefore, this subcategory of datasets may serve as general testbed for the detection of malicious insider threat.

\textbf{CERT Datasets}. 
CERT, along with other partners, has generated a collection of synthetic insider threat datasets. 
The approach employed for the dataset generation is described in~\cite{2013-Glasser}; the datasets were generated using scenarios containing traitor instances, as well as other scenarios involving masquerade activities. 
The logs collected contain logon data, browsing history, file access logs, emails, device usage, psychometric information, and LDAP data.

\textbf{TWOS Dataset.}
The TWOS dataset has been collected by Harilal et al.~\citeyear{2017-twos-mist} as the outcome of a multi-player game designed to reproduce interactions in real companies while stimulating existence of masqueraders and traitors. 
The game involved 24 users, organized into 6 teams that played for one week.
Masquerade sessions were performed by ``\textit{temporarily}'' malicious users, who, once in a while, received credentials of other users (victims) and were able to take control over victim's machines for a period of $90$ minutes. 
Traitor sessions were collected when a few participants were fired from their original team.
The dataset consists of miscellaneous data types such as a mouse, keyboard, network, and host monitor logs of system calls.
Furthermore, the authors presented applicable state-of-the-art features and demonstrated the potential use of the TWOS dataset in multiple areas of cyber-security that relate to the insider threat area, such as authorship verification and identification, continuous authentication, and sentiment analysis~\cite{jowua-TWOS}.

\subsubsection{\textbf{Substituted Masqueraders from Benign Users}}
In this category of datasets, data considered as malicious has been explicitly substituted by legitimate data that was never seen before (e.g., data from other users). 
We created a dedicated category of such datasets, although these dataset can be viewed as related to the authentication problem -- \textit{does an input sample belong to the particular user?} 
Note that previous research has indicated that such datasets are less suitable for testing masquerader detection solutions than \textit{masquerader-based} datasets~\cite{2008-Salem}.

\textbf{Schonlau Dataset}. 
The Schonlau dataset (a.k.a. SEA dataset) was introduced by Schonlau et al.~\citeyear{2001-Schonlau} and contains around 15,000 Unix commands per user; the dataset was generated by 50 individuals who had various roles inside of an organization. 
In this dataset, masqueraders are approximated by randomly blending the data collected from unknown users (i.e., users not among the previously mentioned 50 users), and thus the data used in masquerade sessions does not contain any malicious intent. 
Maxion and Townsend~\citeyear{2002-Maxion} showed that the Schonlau dataset is not appropriate for the detection of masqueraders because of the following:  
1) its data was collected during different time periods for  each user; 
2) each user performed a different number of login sessions;
3) all Unix commands were captured in the order of their termination, instead of the order of their commencement.
Nevertheless, according to the findings of Salem et al.~\citeyear{2008-Salem}, until 2008, SEA served as a common benchmark dataset, leading to a large amount of research aimed at the detection of substituted masqueraders in Unix commands. 

\subsubsection{\textbf{Authentication/Identification-Based Datasets}}
This category of datasets can be used for the purpose of identification or authentication of any user, regardless of his/her intent, although benign intent is implicitly assumed. 
Therefore, this sort of dataset may be used for addressing the identification and authentication questions: 
1) \textit{which user does the input sample belong to?} and 
2) \textit{does an input sample belong to the particular user?} 
The following datasets in this category were used in various insider threat detection works.

\textbf{Greenberg's Dataset}. 
Greenberg's dataset~\cite{1988-Greenberg} is the first known collection of authentication-based data. 
The author collected a dataset comprised of full command-line entries (including arguments and timestamps) from 168 Unix users of the \textit{csh} shell. 
The dataset maintains anonymity of the users, while it retains the semantical information of the executed commands.
The dataset is split into four groups according to the level of user knowledge and skills, and consists of 52 scientists with programming skills, 55 novices, 36 advanced users, and 25 non-technical users, respectively.
The researchers that use Greenberg's dataset usually adapt two different configurations: \textit{plain} and \textit{enriched}.
The plain configuration provides only commands and aliases, and therefore is equivalent to the configuration of the SEA dataset. 
In contrast, the enriched configuration contains commands/aliases together with their full arguments.

\textbf{Purdue University Dataset}. 
The Purdue University (PU) dataset was introduced by Lane and Brodley in~\citeyear{lane1997application}. 
The PU dataset consists of eight subsets of preprocessed Unix command data that were taken from the \textit{tcsh} shell histories of eight computer users at Purdue University during a two year period (initially it contained data from only four users). 
Commands included in this dataset are enriched, and thus contain command names, arguments, and options; nevertheless, filenames are omitted.

\textbf{MITRE OWL Dataset}. 
The Organization-Wide Learning (OWL) dataset from MITRE was designed for collecting application usage statistics in order to provide individual feedback and tutoring to users of an organization~\cite{linton2000owl}, however it was also used for the analysis of human interactions with GUI-based applications for the purpose of user authentication~\cite{2014-ElMasri}. 
During a period of two years (from 1997 to 1998), the data was collected from 24 employees using Microsoft Word on the Macintosh operating system. 
The participants were employees of an artificial intelligence group, researchers, and technical and support staff. 

\section{Analysis of Incidents} \label{Sec:Analysis-of-Incidents}
\begin{figure}[b]
	\centering
	\vspace{-0.3cm}
	\includegraphics[scale=0.45]{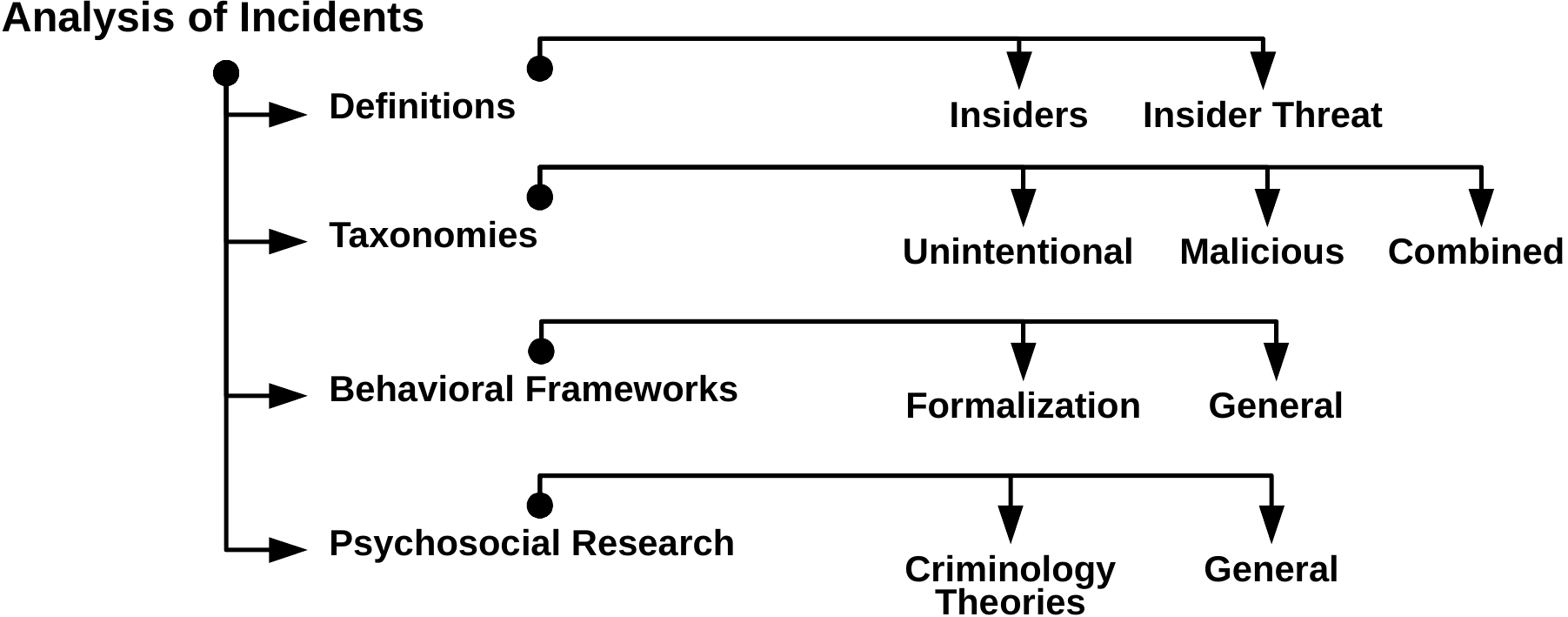}
	\vspace{-0.2cm}
	\caption{Detailed categorization of the \textit{Analysis of Incidents} category\label{fig:categorization-analysis}}
	\vspace{-0.4cm}
\end{figure}

In this section we aim to generalize all related aspects and behaviors of a malicious insider before, during, and after conducting an incident.
We present the research that deals with the analysis of incidents by considering the insider's lifecycle, observed indicators, and the critical pathway. 
First, we consider this mainly from the security perspective and describe behavioral frameworks of insider incidents, and this is followed by consideration of psychological and social theory works. 
Later we will see that significant portion of the research contributions related to the analysis of the incidents serves as a basis for defense solutions (see Section~\ref{Sec:Defense}).
Although definitions and taxonomies are also part of this category, these were discussed in Section~\ref{sec:Taxonomies}, and thus we will not include them here. 
See Figure~\ref{fig:categorization-analysis} for an overview of this category.

\subsection{Behavioral Frameworks}
\begin{figure}[t]
	\vspace{-0.2cm}
	\centering
	\includegraphics[scale=0.41]{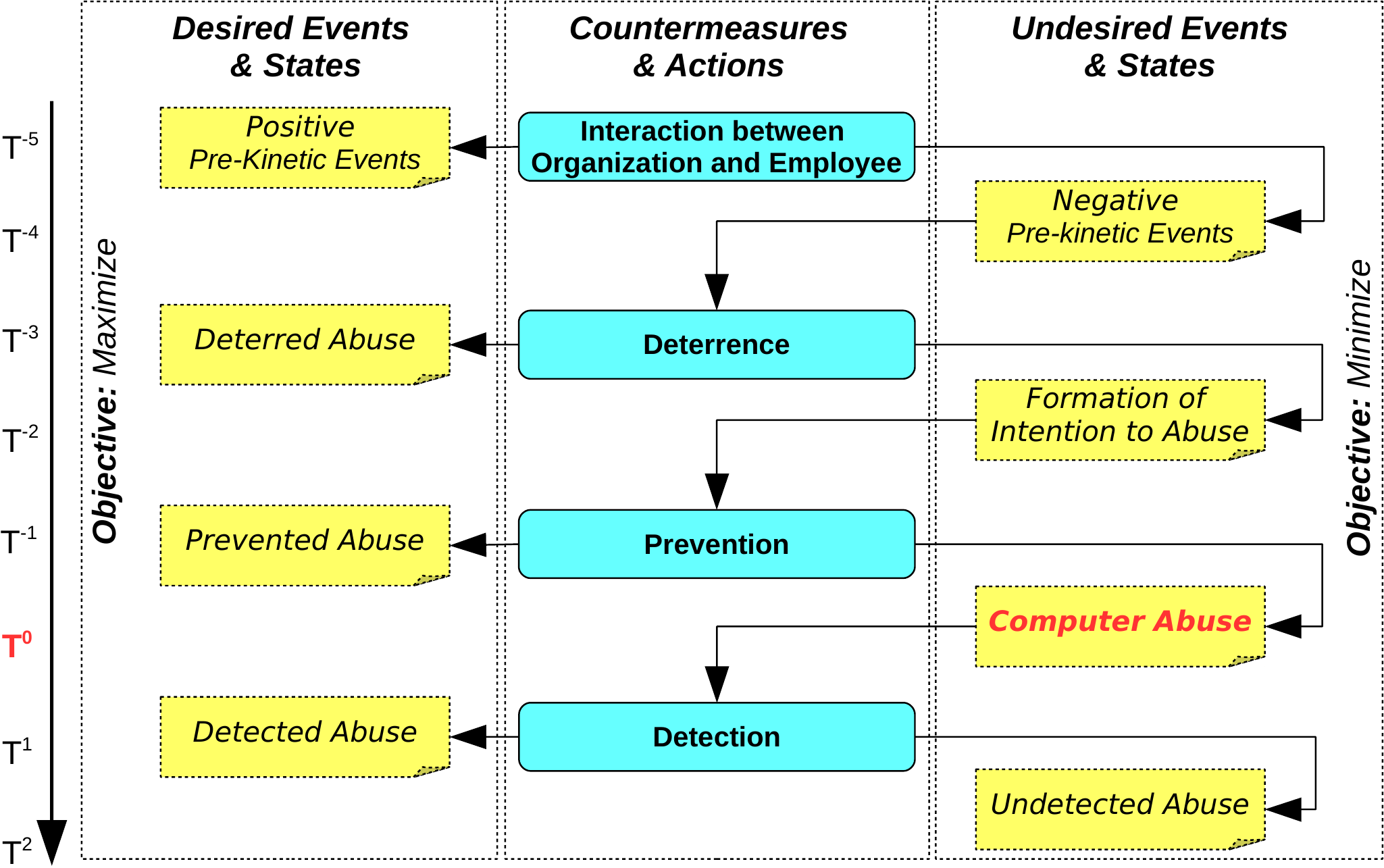}
	\vspace{-0.2cm}
	\caption{Security action cycle (based on~\cite{1998-Straub} and~\cite{2013-Willison}). \label{MI-security-action-lifecycle}}
	\vspace{-0.6cm}
\end{figure}

The security action cycle for inside abuse is a general way in which inside attackers can be modeled in time, which was presented by Straub and Welke~\citeyear{1998-Straub} and later extended by Willison and Warkentin~\citeyear{2013-Willison} who enriched it with pre-kinetic events, which represent temporal antecedents of an act of abuse and occur prior to the deterrence stage. 
This framework is depicted in Figure~\ref{MI-security-action-lifecycle}, in which the vertical axis represents the discrete time relative to the occurrence of computer abuse. 
The middle column represents \textit{Actions \& Countermeasures} that an organization may take in order to address (potential) insider abuse at various stages of its lifetime. 
The successful outcome of such actions and countermeasures appears in the left column (\textit{Desired Events \& States}), while the negative outcome appears in the right column (\textit{Undesired Events \& States}). 
Therefore, the objective of an organization should be to perform Actions \& Countermeasures that maximize Desired Events \& States, and simultaneously minimize Undesired Events \& States.
Like~\cite{2013-Willison}, the remedies category and deterrence feedback loops are not included in the figure for the sake of clarity. 
Working from the top to the bottom of the figure, interaction between the organization and the employee is modeled; pre-kinetic events can be perceived by the employee. 
Positive pre-kinetic events include a friendly and fair working environment, while negative events include organizational injustice, disgruntlement, justification of committing an incident, etc. 
Furthermore, negative pre-kinetic events form negative attitudes that may lead to the intention to perform computer misuse, however this might be deterred by policies, training, or other means. 
If deterrence attempts are insufficient, a misuse attempt is executed, and then it is the turn of prevention controls and implemented policies.
If prevention controls do not stop the misuse act, the act is successfully perpetrated. 
At this point, the organization has a chance of detecting it by technical controls or social means. 

The following include several examples of the behavioral framework subcategory.
Wood~\citeyear{2000-Wood} contributes to this area by elaborating the attributes of malicious insiders: \textit{knowledge, access, privilege, risk, tactics, skills, motivation, and process.}
Andersen et al.~\citeyear{2004-Andersen} presents the system dynamics scheme of the malicious insider threat, based on six case studies involving disgruntlement and financial gain as the motivation factors; this research formed the summary of the 2nd Workshop on System Dynamics Modeling for Information Security.
Based on 10 case studies in the national critical infrastructure industry,  Shaw and Fischer~\citeyear{2005-Shaw} outlined a conceptual framework describing events on the critical pathway of malicious insider attacks, including: 1) \textit{personal/professional stressors,} 2) \textit{emotional and maladaptive behavioral reactions,} 3) \textit{results in official attention,} 4) \textit{ineffective intervention,} and 5) \textit{attack.}
Pfleeger et al.~\citeyear{2010-Pfleeger} proposed a conceptual framework for modeling any insider threat, which is aimed at the risk factors and based on four components and their interactions: \textit{organization, environment, system, and individual.}
Claycomb et al.~\citeyear{2012-Claycomb-2} performed a chronological examination of 15 sabotage case studies, and as a result, the authors identified six consecutive common events: 1) \textit{tipping point} (first observed disgruntlement event), 2) \textit{malicious act} (e.g., installing logic bomb), 3) \textit{occurrence of the attack,} 4) \textit{detection of the attack,} 5) \textit{end of the attack,} and 6) \textit{action on insider} (response).
Nurse et al.~\citeyear{2014-Nurse-2} proposed a conceptual framework for characterizing any form of insider attack, constructed by using grounded theory on 80 case studies.
The framework models consecutively connected aspects of an insider incident: 1) \textit{catalyst,} 2) \textit{actor characteristics,} 3) \textit{attack characteristics,} and 4) \textit{organization characteristics.}
Farahmand and Spafford~\citeyear{2013-Farahmand} investigated accepted models of risk-taking behaviors in the insider threat field and then introduced ordinal scales to a selected model that represents perceived risk as a function of \textit{consequence} and \textit{understanding.} 
Maasberg et al.~\citeyear{2015-Maasberg} proposed a theoretic insider threat model based on the MOC (motive, opportunity, capability) concept, the theory of planned behavior, and dark triad personality traits.
CERT's MERIT project contains, among other parts, system dynamics models whose purpose is to reveal patterns in the course of insider threat cases over time.
Particular profiles of insider threats with corresponding system dynamics models are addressed in research as follows: IT sabotage in~\cite{2008-Moore}, IP theft in~\cite{2011-Moore}, fraud involving financial services in~\cite{2012-Cummings}, and IT sabotage and espionage in~\cite{2006-Band}.

\paragraph{\textbf{Formalization Frameworks.}}
Probst et al.~\citeyear{2006-Probst} proposed a formal model for static analysis of insiders based on two parts: 1) ``a high-level system model based on graphs,'' and 2) ``a process calculus called acKlaim,'' which extends $\mu$Klaim calculus with access control features.
Dimkov et al.~\citeyear{2010-Dimkov} presented PORTUNES, a framework that represents malicious scenarios present in the digital, social, and physical domains; the authors designed and applied abstract language based on Klaim calculus.
Chen et al.~\citeyear{2015-Chen} formally modeled malicious insider threat at micro and macro levels that are represented by an \textit{intentional analysis} and a \textit{behavioral analysis}, respectively. 
The intentional analysis models likelihood of an employee to be an insider threat by Bayesian network, and once an employee intends to attack, the behavioral analysis models the probability of success using Markov decision processes, considering the formal description of an environment in acKlaim calculus.	
Kamm{\"u}ller et al.~\citeyear{2016-Kammuller} formally modeled both malicious and inadvertent insider threats in the IoT using attacks trees and the higher order logic interactive theorem prover, Isabelle.  

\subsection{\textbf{Psychological and Social Theory}}
Prior psychological and social research found correlations between the insider incidents committed and personality factors, current emotional states, predispositions to malicious behavior, and mental disorders~\cite{2013-Greitzer-2},~\cite{2006-Shaw},~\cite{2012-Cappelli},~\cite{2006-Cole},~\cite{ambrose2002sabotage}.
The following works examine psychological or sociological aspects of insider threat.

Shaw et al.~\citeyear{1998-Shaw} focused on critical information technology insiders (such as system administrators, programmers, network professionals) and explored personal and cultural vulnerabilities consisting of: 1) \textit{``introversion,''} 2)  \textit{``dependency on computers,''} 3) \textit{``personal and social frustrations,''} 4) \textit{``ethical  flexibility,''} 5) \textit{``low loyalty,''} 6) \textit{``feeling of entitlement,''} and 7) \textit{``lack of empathy.''}
Potential indicators for the prediction of malicious insider attacks were proposed by Schultz~\citeyear{2002-Schultz} and consist of ``\textit{deliberate markers, meaningful errors, preparatory behaviors, correlated usage patterns, verbal behavior, and personality traits.}''
Leach~\citeyear{2003-Leach} analyzed several factors influencing employees' secure behavior, while emphasizing that three of them improve such secure behavior: 1) \textit{the behavior of other personnel, especially managers,} 2) \textit{the employee's ``common sense of security'' and decision-making experiences,} and 3) \textit{the employee's ``psychological contract with the company.''}		
Ho~\citeyear{2008-Ho} investigated changes in trustworthiness as indicators of insider threat, with trustworthiness defined as ``the degree of correspondence between communicated intentions and behavioral outcomes that are observed over time.''
Farahmand and Spafford~\citeyear{2009-Farahmand} investigated well-known models of risk perception in relation to the benefits of performing malicious actions by insiders.
The authors argue that cognitive understanding and potential consequences are the most significant features influencing a risk perceived by insiders.
Willison and Warkentin~\citeyear{2009-Willison} addressed how perceptions of fairness are formed inside the workplace environment; denoted as \textit{Organizational Justice}, which consists of four constructs: a) \textit{distributive} (e.g., differences in the rewards of employees with the same responsibilities), b) \textit{procedural} (e.g., in disputes), c) \textit{interpersonal} (e.g., managers treat their subordinates respectfully and with dignity), and d) \textit{informational justice} (e.g., discussion of demotion decisions).
Later, Willison and Warkentin~\citeyear{2013-Willison} reviewed existing \textit{neutralization techniques} of insiders that rationalize their malicious behavior.
The level of trust and its consequences to insider threat as part of the risk analysis process are examined in~\cite{2010-Probst-2}.
Ethical and social issues of insider threat monitoring are examined in~\cite{2010-Greitzer-2}.
Martinez-Moyano at al.~\citeyear{2011-Martinez}	focused on behavior-based detection of terrorists, considering the level of technical expertise, history of suspicious activities, and intensity of religious radicalism.
An investigation of causal reasoning of insiders after the implementation of new security measures was performed by Posey et al.~\citeyear{2011-Posey}.
The relative risk of insider attack based on \textit{dynamic environmental stressors, static/dynamic personal characteristics, capability, and counterproductive behavior} was modeled with Bayesian networks in~\cite{2013-Axelrad}.
Based on the analysis of the game they conducted, Ho et al.~\citeyear{2016-Ho} supported the hypothesis that language incentives in group communication change significantly once an insider has turned to the malicious side.

\paragraph{\textbf{Criminology Theories}}
Behavior pertaining to insider misuse, which can be also considered as a type of misbehavior in the workplace, has been studied in criminology research as well. 
Related studies aim at analyzing insider misuse based on various criminology theories. 
Theoharidou et al.~\citeyear{2005-Theoharidou} discussed major criminology theories with regards to ISO 17799 (a standard in IS security management); they focus on theories applicable to insider misuse -- general deterrence theory (GDT), social learning theory (SLT), social bond theory (SBT), theory of planned behavior (TPB), and theory of situational crime prevention (SCP).
Lee and Lee~\citeyear{2002-Lee} adopted the framework of TPB for computer abuse within organizations and assessed the influences of GDT, SBT, and SLT on TPB.
Willision and Siponen~\citeyear{2009-Willison-2} showed how SCP techniques can be utilized with the universal script framework of insider threat; they demonstrated its use on an example of fraud.

\section{Simulation Research}\label{Sec:Simulations}
\begin{figure}[b]
	\vspace{-0.4cm}
	\centering
	\includegraphics[scale=0.47]{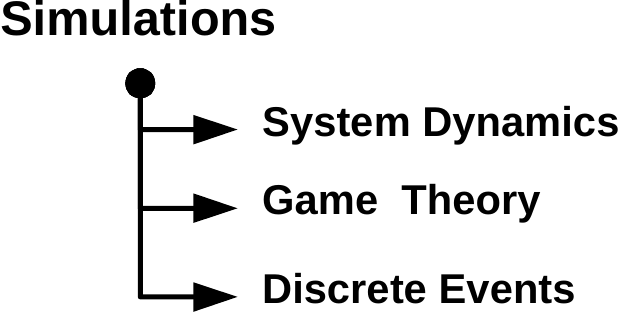}
	\vspace{-0.3cm}
	\caption{Detailed categorization of the \textit{Simulations} category\label{fig:categorization-simulations}}
	\vspace{-0.4cm}
\end{figure}

This section includes papers related to the modeling and simulation domain, as well as some game theory approaches used for the purpose of simulation.
According to Banks~\citeyear{banks1998handbook}, ``simulation is the imitation of a real system's operation over time,'' and it contains creation of \textit{``an artificial history''} of a system's model and ``drawing inferences concerning the operational characteristics of the real system that is represented,'' based on that artificial history.
All studies in this section contain experiments based on the execution of programmed models as simulations.
We have identified three subcategories: 1) \textit{approaches working with discrete events,} 2) \textit{system dynamics approaches,} and 3) \textit{game theory approaches.}
See Figure~\ref{fig:categorization-simulations} for an overview of this category.

\vspace{-0.2cm}
\paragraph{\textbf{Discrete Events}}
According to Banks~\cite{banks1998handbook}, the goal of discrete event simulation is ``to portray the activities in which entities of the system engage, and thereby learn something about the system's dynamic behavior'';
this is accomplished ``by defining the states of the system and constructing activities that move it from state to state.''
The start and end of each activity is represented by events and the state of the simulated system remains unchanged between two consecutive events.
The following simulation studies belong to this subcategory.

Althebyan and Panda~\citeyear{2008-Althebyan-2} utilized this type of simulation as a performance evaluation technique for an insider threat risk assessment model. 
In a similar vein, Alghamdi et al.~\citeyear{2006-Alghamdi} evaluated the performance of proposed multi-entity Bayesian networks, while using two networks: a \textit{generative network} for simulating the observations and an \textit{inference network} for inferring whether a user is an insider threat or not. 
For the purpose of performance evaluation, a discrete time Markov chain was used as a database transaction simulator in~\cite{2013-Panigrahi}.
Simulation in OMNet++ for the purpose of data generation was utilized in~\cite{2009-Eberle-2}; the data had a graph-based representation and was related to business transactions and processes inside of organizations.
Using colored Petri nets, Nasr and Varjani~\citeyear{2014-Nasr} performed threshold-based anomaly detection of simulated malicious insiders in a supervisory control and data acquisition (SCADA) system.

\vspace{-0.2cm}
\paragraph{\textbf{System Dynamics}}
According to Richardson et al.~\citeyear{Richardson2001}, system dynamics is ``a computer-aided approach to policy analysis and design that applies to dynamic problems arising in complex social, managerial, economic, or ecological systems,'' meaning that any dynamic system is represented by dependencies among its components, mutual interactions, feedback loops, and circular causalities. 
Thus, system dynamics enables us to analyze complex systems containing various causalities among their components, and more specifically, it allows us ``to investigate the effect of changes in one variable on the other variables over time''~\cite{2008-Martinez}.
Note that system dynamics simulation may run either in continuous or discrete time, and the state of a simulated system can be evaluated at any point in time. 
The following studies belong to this domain.

Simulations aimed at the inspection of four implementation levels of formal security controls (\textit{absent, poor, normal, and high}) and their impact on the perception of security by management, security level, and incident cost were performed in~\cite{2003-Melara}.
In the paper, the authors simulated the real sabotage case of Tim Lloyd/Omega. 
Rich et al.~\citeyear{2005-Rich} applied system dynamics simulation in long-term fraud detection by evaluating the alignment training of ``information workers'' and ``security officers'' compared to the base run.
During the base run, information workers adjusted their decision thresholds solely on their own, while in the alignment training, some experiences of the security officers were transferred to the information workers. 
Martinez-Moyano et al.~\citeyear{2008-Martinez} inspected a similar fraud detection scenario, however they evaluated three policy strategies against the base run: \textit{perfect information}, \textit{consistency training}, and \textit{alignment training}.
In the base run, the firm guarded itself against inside attackers unsuccessfully in comparison to: 
1) perfect information, where judges (security officers and information workers) did not change their judging strategies but used better information cues; 
2) consistency training in which the defenders were trained to become better judges who responded consistently; and 
3) alignment training that used the same tactics as~\cite{2005-Rich}, enriched by training the security officers who should ensure adherence to security standards by all employees.
In the latter paper, Martinez-Moyano et al.~\citeyear{2011-Martinez} also focused on the convergence of the decision thresholds of security officers for the detection of terrorist activity.

\vspace{-0.32cm}
\paragraph{\textbf{Game Theory}}
Myerson~\citeyear{myerson1997game} defines game theory as ``the study of mathematical models of conflicts and cooperation between intelligent and rational decision-makers.''
Game theory consists of methods that can be used for the analysis of problems in which two or more entities make decisions influencing one another's benefit.
The term \textit{game} represents a social situation that involves two or more entities who are called \textit{players}.	
The following studies contribute to this domain.

Using Nash equilibrium, Liu et al.~\citeyear{2008-Liu} proposed a two-player stochastic game between a system administrator and an insider who perpetrates a fraud.   
Aimed at anomaly detection of insider threat, Zhang et al.~\citeyear{2010-Zhang} proposed a few algorithms for ``the establishment of the defender's reputation,'' which resulted in improvement of the trade-off between false positives and true positives.
Kantzavelou and Katsikas~\citeyear{2010-Kantzavelou} used quantal response equilibrium (QRE), which adjusts players' preferences, and the Von Neumanne-Morgenstern utility function that assigns numbers reflecting players' preferences.
In a comparison to Nash equilibrium, QRE was capable of capturing players' bounded rationality, and thus enabled the players to select even not necessarily the best action.
Tang et al.~\citeyear{2011-Tang} applied dynamic Bayesian networks consisting of various user-behavioral variables as part of a defender/insider game that used QRE.

\vspace{-0.2cm}
\section{Defense Solutions}\label{Sec:Defense}
\begin{figure}
	\centering
	\vspace{-0.4cm}
	\includegraphics[scale=0.44]{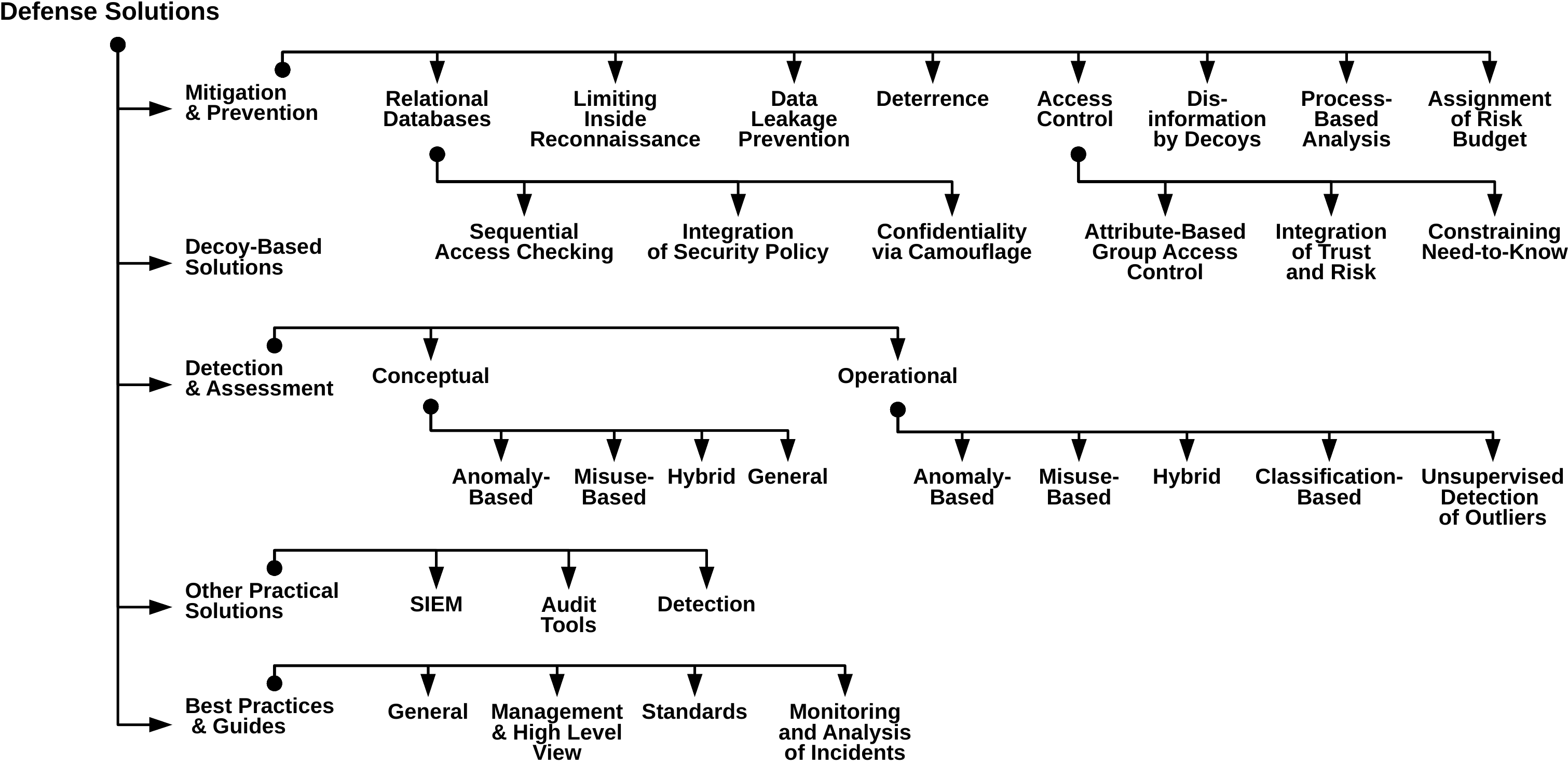}
	\vspace{-0.2cm}
	\caption{Detailed categorization of the \textit{Defense Solutions} category\label{fig:categorization-defense}}
	\vspace{-0.3cm}
\end{figure}

This section categorizes and briefly describes defense solutions for the insider threat problem.
We emphasize that many of the solutions presented may be considered as general IT security mechanisms, and we survey them here because of the relevancy and the importance to insider threats.
First,  means-based categorization of \textbf{mitigation/prevention approaches} is presented, and this is followed by intrusion-detection-derived categorization of insider threat \textbf{assessment and detection} research.
As complementary defense solutions, we identified \textbf{best practices and guidelines}, \textbf{decoy-based solutions}, and \textbf{other practical solutions}.
However, due to space constraints, we refer the reader to appendices~\ref{appendix:best-practices}, \ref{appendix:decoys}, \ref{appendix:other-solutions} for descriptions and examples of these subcategories.
A detailed overview of the defense solutions category is depicted in Figure~\ref{fig:categorization-defense}.
It is important to state that because of the broad nature of insider threats, some of the defense solutions discussed span across other topics that are out of the scope of this survey.
More specifically, we refer the reader to the surveys of Shabtai et al.~\citeyear{2012-Shabtai} and Alneyadi et al.~\citeyear{2016-Alneyadi} in the area of data leakage, and recommend the works of Lazouski et al.~\citeyear{Lazouski201081}, Servos and Osborn~\citeyear{servos2017abac} in the area of access control.
Additionally, some related works have focused on the detection and mitigation of specific kinds of fraud attacks, such as financial fraud, telecommunication fraud, Internet marketing fraud, or insurance fraud, which may be perpetrated by outsiders as well as by insiders (depending on the type of fraud).
The defense solutions against fraud perpetrators are specific to the type of fraud.
For details about the defense techniques used against various kinds of fraud, we refer the reader to the surveys of Abdallah et al.~\citeyear{abdallah2016fraud} and Edge et al.~\citeyear{edge2009survey}.
In this section, we will only include the works that have a significant intersection of insider threat with other topics.

\subsection{\textbf{Mitigation and Prevention}}\label{sec:mitigation-and-prevention}
Studies dealing with the mitigation and prevention of insider threat are divided into several categories according to the type of prevention technique or specific application domain.

\subsubsection{\textbf{Deterrence}}
Vance et al.~\citeyear{2012-Vance} presented several subconstructs of accountability as deterrent factors in policy access violation and conducted a factorial survey to support their assumptions.
The authors showed that convenient features of the user interface of an IS contribute to awareness about accountability, which as a result, cause deterrence of policy violations.

\subsubsection{\textbf{Data Leakage Prevention}}
Wu et al.~\citeyear{2011-Wu} proposed the concept of ``active data leakage prevention'' employing an encrypted secure data container (SDC) and ensuring that data is accessed by authorized users in a trusted environment, while data access and manipulation respect expected patterns.
Johnson et al.~\citeyear{2016-Johnson} proposed SimpleFlow, ``an information flow-based access control system'' that delays action ``until a process tries to write confidential data outside of the local network.''
Pramanik et al.~\citeyear{2004-Pramanik} proposed context-based security policies that prevent illegal information flow among documents; this is based on the idea of prohibiting modification of a file when another inappropriate file is open. 
Bertino and Ghinita~\citeyear{2011-Bertino} proposed an approach for the prevention of data exfiltration, which uses ``provenance tracking'' through the watermarking and ``the confine \& mark method.''
The confine step is achieved using network segmentation and virtualization, whereas the mark step is handled by tokens.

\subsubsection{\textbf{Process-Based Analysis}}
Bishop et al.~\citeyear{2014-Bishop} proposed a process-based approach for the identification and elimination of places where data exfiltration and sabotage can be perpetrated, by applying fault tree analysis and finite-state verification techniques.

\subsubsection{\textbf{Assignment of Risk Budget}}
Liu et al.~\citeyear{2009-Liu-2} proposed a mitigation technique for unintentional insiders by assigning each user a risk budget that specifies the maximum amount of accumulated risk an employee can cause during the execution of a task; employees may be rewarded for staying within their budget or punished for depleting the budget.

\subsubsection{\textbf{Disinformation by Decoys}}
Stolfo et al.~\citeyear{2012-Stolfo} proposed a combination of behavioral monitoring and offensive decoy technology for the prevention of unauthorized access by masqueraders in cloud-based environments. 
When unauthorized access is suspected, the system triggers a ``disinformation attack'' that confuses a potential attacker by a large volume of decoy data.

\subsubsection{\textbf{Relational Databases}\vspace{-0.25cm}}
\paragraph{\textbf{Sequential Access Checking.}}
Chagarlamudi et al.~\citeyear{2009-Chagarlamudi} proposed a concept that prevents execution of malicious users' tasks by checking the partial order of database transactions assigned to each particular application's task against predefined Petri network models. 
Yaseen and Panda~\citeyear{2011-Yaseen} proposed an insider mitigation strategy for sequential access to particular data in the domain of relational databases, using threat prediction graphs.

\textbf{\textit{Confidentiality via Camouflage.}}
Gopal et al.~\citeyear{2002-Gopal} proposed a concept denoted as confidentiality via camouflage (CVC) that performs deterministically numerical interval-based responses of ad hoc queries to relational databases, while preserving confidentiality. 
Garfinkel et al.~\citeyear{2002-Garfinkel} proposed an extension of this approach for binary fields and later proposed an improvement for CVC, aimed at insiders that have knowledge of some data in the confidential fields~\citeyear{2006-Garfinkel}.

\textbf{\textit{Integration of Security Policy.}}
Considering various role access, Jabbour and Menasc{\'e}~\citeyear{2009-Jabbour} proposed a concept for protecting a database environment against insider threat by integrating a security policy mechanism as an inseparable part of the protected system.

\subsubsection{\textbf{Access Control}\vspace{-0.25cm}}

\paragraph{\textbf{Attribute-Based Group Access Control}}
Bishop et al.~\citeyear{2008-Bishop-2} proposed attribute-based group access control (ABGAC) utilized with Carlson's unifying policy hierarchy as a concept for mitigation of malicious and inadvertent insider threats.

\textbf{\textit{Integration of Trust and Risk.}}
Crampton and Huth~\citeyear{2010-Crampton} extended the concept of access control by providing support for awareness of trustworthiness and risk.
In a similar vein, Baracaldo and Joshi~\citeyear{2012-Baracaldo} proposed an extension of the role-based access control (RBAC) concept by applying a risk assessment of roles and the system's trust in its users.

\textbf{\textit{Constraining Need-to-Know.}}
Focused on the document access problem, Aleman-Meza et al.~\citeyear{2005-Aleman} presented an ontological approach based on measuring the distance of access requested documents from the domain of the insider's need-to-know. 
Desmedt and Shaghaghi~\citeyear{2016-Desmedt} proposed the concept of ``function-based access control (FBAC),'' inspired by functional encryption, using operations that access only particular parts of a document (i.e., atoms). 
Shalev et al.~\citeyear{2016-Shalev} proposed a Linux container-based solution for isolating the system administrators from resources irrelevant to their current ticket's task, while enabling them to obtain additional permissions when approved by the permission broker. 
As part of the their work, Eom et al.~\citeyear{2011-Eom} proposed the concept of a misuse monitor that can control access to resources by matching the actual processing pattern to the expected processing patterns. 
Althebyan and Panda~\citeyear{2007-Althebyan} proposed a conceptual model for the prevention of data exfiltration by insiders, which is based on an insider's knowledge base and dependency graphs among documents. 
The proposed model prevents access when the insider's knowledge exceeds an access limit for a particular document cluster. 
The concept of capability acquisition graphs was presented by Mathew et al.~\citeyear{2008-Mathew}, who proposed performing periodical evaluations of the privileges accumulated by users with respect to critical information assets.

\subsubsection{\textbf{Limiting Inside Reconnaissance}}
Achleitner et al.~\citeyear{2016-Achleitner} proposed a network deception system based on software-defined networking (SDN), which defends against reconnaissance (e.g., advanced persistent threat -- APT) conducted by insider adversaries.
The system simulates virtual network topologies that are able to thwart network reconnaissance by delaying the scanning results of attackers, and moreover, the system deploys honeypots that enable the identification of attackers. 
Markham and Payne~\citeyear{2001-Markham} proposed a second generation hardware firewall called network edge security (NES), which is placed on the host's NIC and contains dedicated memory and CPU to which the host does not have access. 
NES can only be accessed remotely. 

\subsection{\textbf{Detection and Threat Assessment Approaches}}\label{subsec:detection-and-assessment}
Works in this section are organized into two subcategories: \textit{conceptual} and \textit{operational}. 
A paper is considered conceptual if its contribution is theoretical (i.e., no relevant empirical results are presented),
while it is considered operational if the authors include a proof-of-concept (i.e., evaluate their solution against a relevant dataset). 
These two groups are further subclassified based on criteria derived from intrusion detection, which is described below.
Note that we only include several of the most significant and cited works here; the rest of the studies included in our survey is presented in Appendix~\ref{appendix:detection-extra}.

\subsubsection{\textbf{Conceptual Works}}
Considering two general classes of intrusion detection -- misuse-based and anomaly-based detection -- we have divided the conceptual techniques for the detection of insider threat into four subcategories that have specific characteristics: \textit{anomaly-based}, \textit{misuse-based}, \textit{hybrid}, and \textit{general}. 
Anomaly-based approaches model legitimate behavior as a baseline profile, and with new input, they compute a score that represents the distance from the baseline profile. 
In contrast, misuse-based approaches model malicious behaviors and measure the similarity or conformity of new input with them. 
The anomaly and misuse-based categories can be viewed as one-class techniques, as they model just one type of behavior. 
On the other hand, there are other types of approaches that simultaneously model both kinds of behavior and therefore can be considered two-class techniques; we denote them as hybrid techniques. 
Finally, we include a subcategory for approaches that aim at general reasoning in detection of insider threat, and denote them as general techniques.

\vspace{1mm}
\textbf{\textit{Anomaly-Based.}}
In the relevant literature, anomaly-based conceptual techniques consider a wide range of observables, such as host-based observables, psychosocial indicators, cyber-environmental properties, etc.
Early conceptual works in our database just tended to propose the use of cyber observables.
For example, Magklaras and Furnell~\citeyear{2002-Magklaras} developed ``evaluated potential threat'' (EPT) metrics that model user behavior by features, such as role, file system knowledge, access to critical components, previous intrusion records, and network activity.
Later, these authors proposed a method for the estimation of end user sophistication~\citeyear{2005-Magklaras}, presuming it to be a potential factor influencing the capability of users to commit insider misuse.
Further trends in the insider threat detection research brought \textit{role-based monitoring}, \textit{natural language processing} (NLP), \textit{assessing the environment}, \textit{business process monitoring}, and \textit{psychological indicators}.
Considering \textbf{role-based monitoring}, Ali et al.~\citeyear{2008-Ali} proposed host-based user profiling with policies defined by a role-based trust matrix, where access decisions are made according to user-specific thresholds.
Focusing on the \textbf{database domain}, Bertino and Ghinita~\citeyear{2011-Bertino} proposed to use pattern matching techniques for the detection of data exfiltration anomalies of database users whose baseline profiles are created during normal periods of activity.
Considering \textbf{psychological factors}, Kandias et al.~\citeyear{2010-Kandias} used psychometric tests based on the  MOC model for the computation of a per user threat score. 
\textbf{NLP} is a branch of techniques usually used to infer various psychological or emotional indicators (sentiment analysis), which can be used for other purposes as well.
For example, Raskin et al.~\citeyear{2010-Raskin} addressed an unintended inference by identifying hidden information from social media and conversations.
In terms of \textbf{assessing the environment}, Althebyan and Panda~\citeyear{2007-Althebyan} adapted their previous approach to Bayesian networks (BN), creating the knowledge Bayesian attack graph, which enabled them to estimate risk values for various objects in a system by Bayesian inference. 
Considering \textbf{business process monitoring}, Gritzalis et al.~\citeyear{2014-Gritzalis} focused on detecting performance deviations of users, which were correlated with other indicators from social media and technical controls.

\vspace{1mm}
\textbf{\textit{Misuse-Based.}}
These approaches do not usually occur in the well-known signature matching form, but rather incorporate softer forms of matching represented by similarity measurement. 
Ray and Poolsapassit~\citeyear{2005-Ray} utilized concepts of \textit{attack trees} for online monitoring and assessment of insider threat; their assessment is based on a comparison of the minimal attack tree of a system, which is generated according to the initially specified user's intent, and the user's runtime actions on a system. 
Utilizing active directory services, Bhilare et al.~\citeyear{2009-Bhilare} proposed a \textit{rule-based} concept for the detection of insider threat violating policies in an academic campus environment. 
Agrafiotis et al.~\citeyear{2016-Agrafiotis} proposed \textit{tripwire grammar} capable of capturing abstraction of policies that organizations adopted, as well as signatures of insider misbehaviors. 
As part of this technique alerts are generated when policy is violated or signature of insider misbehavior is matched.

\vspace{1mm}
\textbf{\textit{Hybrid.}}
Approaches simultaneously combining anomaly and misuse-based detection belong to this category. 
Liu et al.~\citeyear{2009-Liu} proposed a conceptual model called \textit{sensitive information dissemination detection} (SIDD) for the detection of insider threats involving the exfiltration of sensitive data to external networks of an organization. 
SIDD is a network device that is placed at the edge of the network and transparently performs three tasks: identification of applications from the payload of network packets, matching of content signatures, and detection of covert channels.\footnote{We consider this work as a conceptual one, as the evaluation was not performed using data clearly related to insider threat.}
Considering observables outside of cyberspace (situation-aware observables), Buford et al.~\citeyear{2008-Buford} described an architecture for insider threat detection based on \textit{belief-desire-intention (BDI) agents} that model behaviors of user roles and malicious insiders and compares these behaviors to the predefined set of plans.
Aimed at insiders in the intelligence community, Park and Ho~\citeyear{2004-Park} proposed the \textit{composite role-based monitoring} (CRBM) approach as an extension to RBAC, having separate role structures for organizations, operating systems, applications, and their mutual mappings. 
In CRBM, a user's behavior is monitored in three separated sessions (OS, application, and organization), and it is compared with expected and unexpected behaviors.

\vspace{1mm}
\textbf{\textit{General.}} 
A general framework of a tiered conceptual detector was presented by Legg et al.~\citeyear{2013-Legg} who designed the framework on the bases of \textit{top/down and bottom/up reasoning in hypothesis trees}, while incorporating three layers: hypothesis, measurement, real world.

\subsubsection{\textbf{Operational Works}}
In this section we briefly survey insider threat detection and assessment approaches that contain a proof-of-concept evaluated on relevant data.
Similar to the previous section, we grouped the approaches into five categories reflecting intrusion detection classification, which is in addition to the previous section enriched by machine learning-based perspective. 
The resulting categories are: \textit{anomaly-based}, \textit{misuse-based}, \textit{hybrid}, \textit{classification-based}, and \textit{unsupervised outlier detection}. 
Here, the difference between hybrid and classification-based approaches is that the former independently merges misuse-based and anomaly-based types, while the latter does that simultaneously using two-class (or multi-class) classifications techniques. 
In contrast to all of the remaining categories, unsupervised outlier detection category does not require labeled training data.

In addition to intrusion detection-derived categorization, we propose using two other categorizations that can also be applied on operational works dealing with the detection and assessment of insider threat: 1) categorization based on \textit{\textbf{the dataset setting used for evaluation}} (in accordance with Section~\ref{Sec:Datasets}), and 2) categorization based on \textit{\textbf{the feature domains}} (in accordance with~\cite{2016-Gheyas}). 
Further information regarding these additional categorizations are present in appendices~\ref{appendix:taxonomy-features} and~\ref{appendix:supplementary-Material}, which also demonstrates their application.

\vspace{1mm}
\textbf{\textit{Anomaly-Based.}}
Similar to the conceptual works, the most widely adopted approach of operational works is anomaly-based detection.
There are works that deal with specific types of cyber observables, as well as studies that combine such cyber observables. 
In a set of works dealing with \textbf{specific cyber observables}, \textbf{Unix/Linux command histories} are the most widespread data source.
The most well-known work in this domain was done by Schonlau et al.~\citeyear{2001-Schonlau} who evaluated six anomaly-based methods (sequence match, uniqueness, Markov models, compression and incremental probability action modeling (IPAM)); they also provided the research community with the SEA dataset (referred to as SEA in this paper). 
Another type of data source used in this field is \textbf{file system interaction}.
As an example, Cami{\~n}a et al.~\citeyear{2014-Camina2}  proposed detection systems for masqueraders utilizing SVM and k-NN as one-class techniques that were evaluated on the WUIL dataset. 
Another example aimed at detection of masqueraders but using graph partitioning is presented by Toffalini et al.~\citeyear{2018-toffalini} who evaluated their approach on WUIL and TWOS datasets.
Considering \textbf{system calls}, Liu et al.~\citeyear{2005-Liu} performed supervised anomaly detection based on k-NN, in which input features were aggregated by n-grams, histograms, and parameters of system calls. 
Companies' \textbf{databases} represent an attractive target for insiders, and therefore some research has addressed this area specifically. 
For example, Panigrahi et al.~\citeyear{2013-Panigrahi} performed user profiling for the detection of suspicious transactions by using the extended Dempster-Shafer theory to combine multiple pieces of evidence, using inter and intra-transactional features. 
The possibilities of user identification also increased with the introduction of \textbf{GUI}. 
For example, Sankaranarayanan et al.~\citeyear{2006-Sankaranarayanan} used IPAM and compared it with a classification-based Na\"{i}ve Bayes approach for the detection of masqueraders in Microsoft Word user behavior.
In contrast to the previous anomaly-based techniques, an example dealing with \textbf{general cyber observables} was presented by Salem and Stolfo~\citeyear{2011-Salem-2} who proposed an approach using one-class SVM and seven features modeling search behavior that is aimed at detecting anomalies in the RUU dataset. 
In addition to finding correlations between insider threat and \textbf{psychosocial observables}, operational kinds of research involving such observables have also increased. 
For example, Brdiczka et al.~\citeyear{2012-Brdiczka} proposed a traitor assessment using Bayesian techniques that combined ``structural anomaly detection from information and social networks with psychological profiling,'' and then evaluated the approach on the World of Warcraft dataset.

\vspace{1mm}
\textbf{\textit{Misuse-Based.}}
Aimed at insiders performing data exfiltration resulting into underlying changes to the integrity of directory services, Claycomb and Shin~\citeyear{2010-Claycomb} proposed \textit{a combination of policy with monitoring}, which leverages the capabilities of directory virtualization. 
Hanley and Montelibano~\citeyear{2011-Hanley} demonstrated the utilization of \textit{signature alerts in the SPLUNK} logging engine for the detection of data exfiltration. 
Aimed at high privileged system users, Sibai and Menasc{\'e}~\citeyear{2011-Sibai} proposed a system for insider threat detection as a network element that is based on \textit{rule-based policies} (in Snort format) defined for different categories of applications; the system executes decryption of network traffic for payload inspection.
For specifying insider misuse signatures, Magklaras and Furnell~\citeyear{2012-Magklaras} designed the \textit{insider threat prediction and specification language} (ITPSL), which has markup features and utilizes logical operators; they evaluated ITPSL's application on a game containing several malicious and accidental scenarios.

\vspace{1mm}
\textbf{\textit{Hybrid.}}
Aimed at the detection of data exfiltration from \textbf{network data}, Maloof and Ste\-phens~\citeyear{2007-Maloof} proposed the ELICIT system, which is based on 76 binary detectors that examine volumetric anomalies, suspicious behaviors, etc.; the outputs of these detectors are passed into a Bayesian networks (BN) to perform threat assessment. 
Considering cyber observables enriched by \textbf{psychosocial indicators}, Legg et al.~\citeyear{2015-Legg} proposed a hybrid approach based on known attacks and policy violations combined with threshold and deviation-based anomalies. 
Considering MOC-based features and using a high level of abstraction, AlGhamdi et al.~\citeyear{2006-Alghamdi} applied a multi-entity BN for the assessment of insider threat and legitimate behavior in the \textbf{document relevance problem}. 
Maybury et al.~\citeyear{2005-Maybury} presented results of the ARDA NRRC (MITRE) workshop for the US intelligence community, where the authors only considered cyber observables for the fusion of three approaches: 1) honeytokens used as Web pages, 2) stealth watch sensors that monitor abnormal network and host activities, and 3) structured top-down analysis modeling pre-attack indicators.

\vspace{1mm}
\textbf{\textit{Classification-Based.}}
Aimed at data exfiltration, Azaria et al.~\citeyear{2014-Azaria} employed SVM and Na\"{i}ve Bayes as part of their BAIT framework that considers \textbf{cyber data} from actions such as transfer (print, copy to thumb drive) and send (by email, HTTP/HTTPS services), while distinguishing between external and internal actions.
As part of their work, Mathew et al.~\citeyear{2010-Mathew} proposed a data-centric approach to role-based masquerader detection in \textbf{relational databases}, which applied supervised techniques such as SVM, J.48 decision tree, and Na\"{i}ve Bayes classifiers.
The detection of masquerades in \textbf{Unix commands} was addressed by Maxion and Townsend~\citeyear{2002-Maxion}, who used a Na\"{i}ve Bayes classifier with an updating scheme, which considered frequency of particular commands for each user.
Inspired by cache memory term locality, Cami\~na et al.~\citeyear{2016-Camina} addressed masqueraders in \textbf{file system access} based on temporal and spatial locality features processed by TreeBagger -- an ensemble of decision tree classifiers from MATLAB. 
Garg et al.~\citeyear{2006-Garg} proposed the detection of masqueraders in the \textbf{GUI} environment by SVM, considering only mouse derived features such as speed, distance, angles, and their statistical properties; evaluation was performed on data collected from three users. 
For the purpose of assessing the \textbf{trustworthiness of entities} such as actors or documents, Mayhew et al.~\citeyear{2015-Mayhew} proposed behavior-based access control (BBAC), which is based on a sequential combination of k-means clustering and SVM. 
Dealing with \textbf{NLP} in comments of YouTube users, Kandias et al.~\citeyear{2013-Kandias-2} employed SVM, logistic regression, and Na\"{i}ve Bayes classifiers in order to predict users with negative/radical political attitudes, assuming these attitudes to be precursors of insider threat.

\vspace{1mm}
\textbf{\textit{Unsupervised Detection of Outliers.}}
Dealing with masqueraders in \textbf{Unix commands}, Lane and Brodley~\citeyear{1998-Lane} presented an online learning system of user behavior that supports the \textit{concept drift}. 
Considering \textbf{general cyber observables}, Senator et al.~\citeyear{2013-Senator} presented the results of the PRODIGAL team, in which several outlier detection methods were designed; 
the evaluation was performed on data collected by SureView from approximately 5,500 real users and red team scenarios injected by CERT. 
As part of their work, Mathew et al.~\citeyear{2010-Mathew} dealt with cluster-based detection of masqueraders in \textbf{relational databases}.
Aimed at indirect \textbf{psychological observables}, Kandias et al.~\citeyear{2013-Kandias-1} detected outliers based on social network graphs containing more than one million Greek Twitter users, while the authors assumed that narcissism/extroversion is a precursor of insider threat. 
Considering aspects of \textbf{mutual interactions of users}, Okolica et al.~\citeyear{2008-Okolica} expanded probabilistic latent semantic indexing (PLSI) by including users in order to extract an individual's interests from emails, while they modeled sharing of interests with individuals' co-workers versus those interests only shared with people external to the organization. 
More general outlier detection was performed by Eberle and Holder~\citeyear{2010-Eberle} who applied graph-based anomaly detection for the identification of traitors, which was evaluated on: 1) the Enron dataset, 2) cell phone calls from VAST grant, and 3) business processes containing fraud scenarios.

\section{Conclusion}\label{Sec:Conclusion}
The objective of this study was to provide a systematization of knowledge in insider threat research, while leveraging existing grounded theory method for rigorous literature review. 
In so doing, we identified four main categories of research and development efforts (see Figure~\ref{research-workflow}):
\begin{inparaenum}
	\item In the \textit{incidents and datasets} category we provided references to several sources of insider threat case studies, as well as categorization and details about related datasets. 
	\item In the \textit{analysis of incidents} category we provided generalization of aspects and behaviors of insider threat and aimed at including research contributions that addressed an insider attack's lifecycle, indicators, and critical pathways, as well as psychosocial point of view.
	\item In the \textit{simulations} category we described research utilizing modeling and simulation approaches for experiments with programmed detection methods or for the purpose of data generation. 
	We identified three groups of approaches: a) approaches working with discrete events, b) system dynamics approaches, and c) game theory approaches.
	\item The \textit{defense solutions} category, which constitutes the majority of the research included in our survey, was divided into four subcategories: a) mitigation and prevention, b) detection and threat assessment, c) best practices and guidelines, d) decoy-based solutions, and e) other practical solutions.
\end{inparaenum}
In addition to the categorization itself, our intention was to illustrate workflow between categories, which followed the direction from incidents to defense solutions.
Special attention was paid to definitions and taxonomies of insider threat -- we proposed a structural taxonomy of insider threat incidents (see Figure~\ref{fig:unification-of-taxonomy}), which is based on existing taxonomies and the 5W1H questions of the information gathering problem. 
The proposed taxonomy contributes to orthogonal classification of incidents and defining the scope of defense solutions employed against them.

Finally, we made the following observations aimed at highlighting directions and challenges for future research:
\begin{compactitem}
	\item As indicated by Salem et al.~\citeyear{2008-Salem}, ``a major challenge of insider threat detection research is the lack of real data'' for assessing defense solutions; this fact has not changed since then.
	Moreover, only a few synthetic datasets contain samples of malicious insider attacks.
	It is important to note that these synthetic datasets are not validated as correctly modeling real environments.
	Therefore, the challenge of absence of datasets still holds, and we encourage researchers to create and share related datasets with the community.
	\item Assuming that an insider is a moving target, we emphasize that detection methods built upon existing datasets may not work in practice. 
	Therefore, we recommend to pursue adversarial classification for improvement of detection approaches as well as for the enrichment of datasets (e.g., using Generative Adversarial Networks~\cite{goodfellow2014generative}).
	\item  New datasets should be also designed and created with collusion attacks and several variations of concept drift in mind; taking such natural phenomena into account would improve the datasets and enable more realistic and challenging testing of detection methods.
	\item There is a trend in the development of detection approaches based on data collected from single-player and multi-player games. 
	The majority of these games are deterministically guided according to scenarios given to the players. 
	The challenge associated with this centers on providing the players of multi-player games with ethical dilemmas, i.e., whether to perform malicious actions for self-benefit or bolster the team and refrain from performing malicious acts (such as in~\cite{2012-Brdiczka,2016-Ho}).
	\item In Table~\ref{tab:detection-by-dataset-and-features} of Appendix~\ref{appendix:supplementary-Material} we see a trend of detection approaches that utilize opportunity-based features and do not make use of motive and capability-based features. 
	The reasons for the absence of motive-based features (attributed to psychosocial observables and decoys) may involve privacy issues or the fact that the underlying datasets do not contain suitable data. 
	In contrast, capability-based features can often be derived from existing data 
	(e.g., similar to Magklaras and Furnell~\citeyear{2005-Magklaras}). 
	Therefore, another challenge is to include this kind of information in datasets as well as the detection approaches working with them.
	\item Social engineering testing or decoy-based solutions may help defend against malicious insider threat and may also generate valuable data for the design of detection methods.
	\item	We observed a trend toward anomaly-based and unsupervised outlier approaches, which can be attributed to class imbalance in datasets and fear of zero-day malicious attacks. 
	However, we believe that a good and robust insider threat defense program should contain a combination of several independent solutions. 
	In the first line of defense, procedural and other best practices involving mitigation and prevention techniques (such as those mentioned in Section~\ref{sec:mitigation-and-prevention}) should be present. 
	In the second line of defense, there should be misuse-based detection that covers existing insider threat scenarios. 
	Finally, in the third line of defense, anomaly-based and unsupervised outlier detection should be deployed. 
	Alerts from the latter two layers should be correlated together in a dedicated view.
	\item	The last recommendation for future research in defense solutions is to specify the MOC-based feature domains of the approach (as presented in Appendix~\ref{appendix:supplementary-Material}) and enumerate the detection scope covered by the defense solution (e.g., using the taxonomy in Figure~\ref{fig:unification-of-taxonomy}).
\end{compactitem}

\bibliographystyle{ACM-Reference-Format}

\bibliography{lit-transformed}

\appendix
\section*{Appendix}

\section{\textbf{Insider Threat from Different Perspectives}}
\label{appendix:perspectives}
The borders of the insider threat problem are difficult to determine, as the insider threat can be part of various definitions pertaining to several research topics that have emerged in different eras of the history of computer systems.
In this section, we demonstrate this issue on four definitions from four disparate areas:
1) \textit{counterproductive workplace behavior}, 2) \textit{intrusion attempts}, 3) \textit{threats to IS security}, and 4) \textit{malicious insider threat involving third parties}.

\subsubsection*{\textbf{Counterproductive Workplace Behavior}}
Counterproductive workplace behavior (CWB) is defined as ``intentional behaviors that are contrary to legitimate organizational interests''~\cite{sackett2002structure}, and thus involve insider threat.
Unlike insider threat, CWB includes, for example, misuse of IT resources and work time for personal business, misuse of sick leave, purposely slow work, engagement in P2P sharing, etc.
Note that CWB inherently only encompasses the internal employees of an organization.

\subsubsection*{\textbf{Intrusion Attempts}}
Anderson~\citeyear{1980-Anderson} defined threats against computer systems (also denoted as intrusion attempts) as the possibility of an intentional unauthorized attempt to:
1) \textit{access information},
2) \textit{modify information}, or
3) \textit{render a system unavailable for other legitimate users}.
In terms of information security, Anderson referred to attempts to violate \textit{confidentiality}, \textit{integrity}, and \textit{availability} of information (i.e., CIA triad).
According to Anderson, threats are further divided into internal and external threats, where internal threats refer to insider threats and contain masqueraders, misfeasors, and clandestine users (see Section~\ref{TAX-Anderson}), each of which is considered an internal employee of an organization. 
Thus, insider threats can be conceptualized as a subset of intrusion threats~\cite{2009-Myers}.

\subsubsection*{\textbf{Threats to IS Security}}
Loch et al.~\citeyear{1992-Loch} proposed four-dimensional categorization of threats to information system security, also based on the CIA concept, which, in contrast to Anderson~\citeyear{1980-Anderson}, also covers unintentional insider threats, natural disasters, and mechanical/electrical failures. 
The dimensions of this categorization, along with their top-down order is as follows:
	1) \textit{internal} and \textit{external},
	2) \textit{human} and \textit{non-human},
	3) \textit{accidental} and \textit{intentional},
	4) \textit{disclosure, modification, destruction}, and \textit{denial of use};
yielding 16 subcategories.

\begin{figure}[!t]
	\centering
	\includegraphics[scale=0.45]{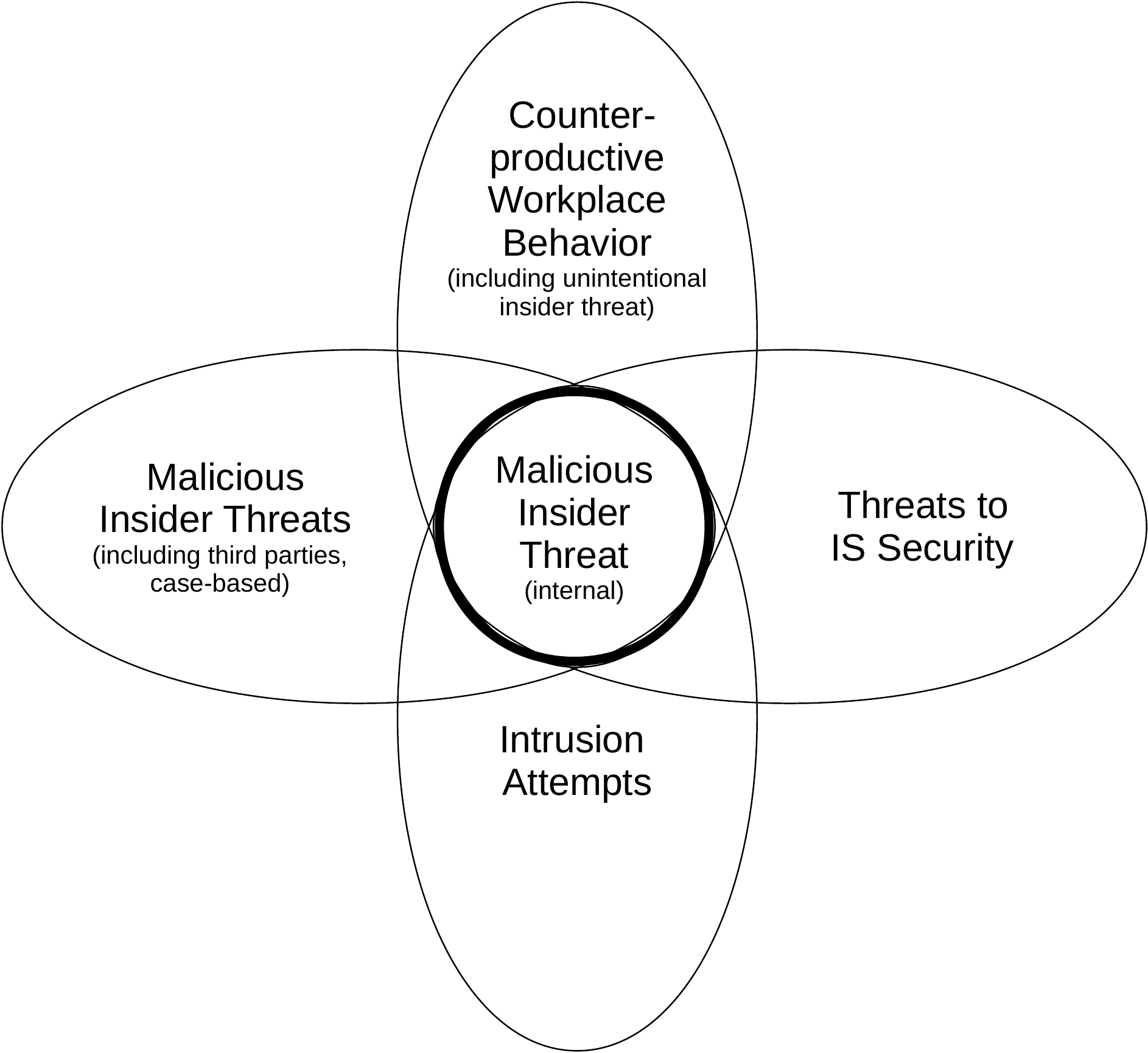}
	\vspace{-0.25cm}
	\caption{Malicious insider threat from different perspectives \label{MI-intersection}}
	\vspace{-0.3cm}
\end{figure}

\subsubsection*{\textbf{Malicious Insiders Involving Third Parties}}
Similar to Anderson's CIA-based definition of threat against computer systems, Cappelli et al.~\citeyear{2012-Cappelli} defined malicious insider threat as ``a current or former employee, contractor or business partner who has or had authorized access to an organization's network, system or data and intentionally exceeded or misused that access in a manner that negatively affected the confidentiality, integrity or availability of the organization's information or information systems.'' 
This definition is based on real case studies from the CERT database. 
In contrast to previous definitions, this definition also explicitly mentions external employees such as contractors or business partners.

If we consider the CIA-based definition of malicious insider threat by Cappelli et al.~\citeyear{2012-Cappelli}, the CIA-based definition of intrusion attempts by Anderson~\citeyear{1980-Anderson}, the CIA-based categorization of threats to information systems from Loch et al.~\citeyear{1992-Loch}, as well as the definition of CWB by Sackett~\citeyear{sackett2002structure}, we can see that there is an intersection of these definitions that is shared by all of them and refers to the internal malicious insider threat for information systems. 
The situation is depicted abstractly in Figure~\ref{MI-intersection}. 
Note that unintentional insider threat can be and aspect of CWB (e.g., not respecting policies), external intrusion attempts (e.g., thwarting by social engineering), and threats to IS security (i.e., inherently covered by the dimensions).

\section{The Application of Proposed Structural Taxonomy of Insider Incidents: Examples}
\label{appendix:structural-categorization-examples}
We present two examples of the application of our proposed structural taxonomy of insider incidents utilizing 5W1H questions. 
The first example is aimed at a case study of a malicious insider incident and the second example contains a case study of an unintentional insider incident.

\paragraph{\textbf{Example of Application for a Malicious Insider Incident}}
Consider a case study involving a malicious insider employed as a security guard (a contractor) in a hospital facility~\cite{2016-Collins-CERTv5}.
In this case, the insider was engaged in the Internet underground, and moreover was the leader of a hacking group.
He worked for the targeted organization at night without any supervision.
The insider's unauthorized activities involved an interaction with the heating, ventilation, and air conditioning (HVAC) computer.
``The HVAC computer was placed in a locked room, but the insider used his security key to obtain physical access to the computer.
The insider remotely accessed the HVAC computer five times over a two-day period, using password cracking programs to attack the organization, and installed a botnet with the intention of conducting a DDoS attack against an unknown external target.
The insider's malicious activities caused the HVAC system to become unstable, which eventually led to a one-hour system outage.
In addition, the insider accessed a nurses station computer, which was connected to all of the victim organization's computers, stored medical records, and patient billing information.''

In this case, the answers to the 5W1H questions of the proposed categorization are as follows: \textbf{What} -- The outcome of the incident is \textit{miscellaneous} as it includes the outage of the HVAC computer and also has some aspects of \textit{fraud}, because the insider accessed private patient information that might be sold on the dark web, and it also involves the unauthorized use of resources for the purpose of a DDoS attack. 
\textbf{How} -- The insider obtained \textit{unauthorized access} to the HVAC computer by using his authorized access to enter the locked room. 
\textbf{Why} -- We consider the insider as \textit{planted}, because it is likely that the reason the leader of a hacking group worked as a security guard was so that he could exploit the physical access provided by this line of work. 
The motivation of the insider is not clear from the description, but it seems to be financial, as providing the botnet may yield money, and there is money to be gained in selling stolen private records. 
Moreover, the motivation could also be considered as \textit{personal} if the insider perpetrated the incident solely because of curiosity. 
\textbf{Who} -- Despite the nature of the insider's position which is not typical for \textit{high-end} insiders, he has some characteristics of this profile, and we can consider his job as a mission. 
The incident belongs to the \textit{inside associate} category, \textit{hacker} subtype, with an \textit{external} job contract and a system role of \textit{none}, since the insider was not granted legitimate access to any system. 
\textbf{Where} -- The inside attack might be detected at the \textit{operating system}, \textit{network}, and \textit{physical } levels (unusual time and activities on the HVAC computer; a password cracker that attacked other machines; and entering the locked room), while the DDoS attack that followed might be detected at the \textit{operating system} and \textit{network} levels (suspicious processes of DDoS application; execution of the DDoS attack).
\textbf{When} -- The two-day duration of the malicious activities makes it \textit{short-term}, and the incident was committed \textit{before} the insider's job was terminated.

\paragraph{\textbf{Example of Application for an Unintentional Insider Incident}}
Consider a case study involving a social engineering attack from~\cite{cert2014unintentional}.
In this case, due to previous incidents of social engineering attacks, employees received training in this area, however some time later some of them were deceived by a phishing email about HR benefits.
The clicking on a vulnerable link caused exploitation of a zero-day vulnerability, followed by download of malicious code.
``Only a few megabytes of encrypted data were stolen, but the organization failed to recognize additional dormant malicious code.
The organization was forced to disconnect Internet access after administrators discovered data being externally siphoned from a server.''

In this case, the answers to the 5W1H questions of the proposed categorization are as follows:
\textbf{What} --
\textit{Data leakage} was the outcome of this incident.
\textbf{How} -- 
\textit{Installed malicious code} through phishing was the type of the attack, while the origin of the attack was an \textit{outsider} who \textit{duped} an insider.
An unintentional violation of policy was inflicted on by using a legitimate access.
\textbf{Why} -- 
Despite the fact that the employees had previous training in order to resist phishing, they succumbed to it.
This could have been the result of \textit{insufficient training} that did not cover highly obfuscated incidents of phishing, as well as \textit{insufficient technical controls} that enabled the attacker to deliver email with a spoofed internal sender address from the outside.
\textbf{Who} -- The insiders belong to the \textit{social engineered} type, and they had an \textit{internal} job contract with no system role mentioned in the description of the incident.
\textbf{Where} -- The phishing part of the attack might be detected at the \textit{application} level (by advanced mechanisms that monitor and profile memory accesses of an application that contains a vulnerability), and the following DDoS attack might be detected at the \textit{operating system} and \textit{network} levels (suspicious processes of the DDoS application; execution of the DDoS attack).
\textbf{When} -- The incident \textit{recurred} in the past.

\section{\textbf{Taxonomy of Detection Features}}
\label{appendix:taxonomy-features}
Gheyas and Abdallah~\citeyear{2016-Gheyas} proposed a taxonomy of insider threat detection features based on the MOC model: 1) \textit{motive}, 2) \textit{opportunity}, and 3) \textit{capability}, which is further divided hierarchically by feature domain. 
Below we present some specific feature domains from these three categories, along with a few examples of features:

	\subsubsection*{\textbf{Motive.}} The motive refers to ``the reason why an insider or group of insiders perpetrate a crime.''
	The examples include:
		\textbf{predisposition to malicious behavior} (e.g., trap-based decoys such as honeypots or honeytokens, information inferred from social networks such as delinquent behavior in the past, swearing or vulgar comments toward law enforcement and authorities);
		\textbf{mental disorders} (e.g., depression, schizophrenia, paranoia, bipolarity);
		\textbf{personality factors} (e.g., neuroticism, narcissism, conscientiousness);
		\textbf{current emotional state} (e.g., anger, fear, stress, surprise, disgust).

	\subsubsection*{\textbf{Opportunity.}} The opportunity features refer to various roles of insiders and the activities they may perform.
	The examples include:
		\textbf{roles} (e.g., system and project roles assigned to a user);
		\textbf{login}  (e.g., logins on a user's PC, logins on other PCs, session duration, login frequency);
		\textbf{file}  (e.g., accessed directories, files which were copied, created, deleted, moved, modified);
		\textbf{database}  (e.g., accessed and modified database items, the number of tables/schemes accessed per time interval);
		\textbf{HTTP}  (e.g., accessed URLs and domains information, encryption of websites, browser information);
		\textbf{removable Devices}  (e.g., device name and type, file operations per time);
		\textbf{email}  (e.g., source and destination, destinations outside of the organization, communication patterns, names and types of attachment);
		\textbf{mobile calls}  (e.g., source and destination, date and time of calls, duration, communication patterns);
		\textbf{printing}  (e.g., name of documents, the number of copies, time-based copy patterns);
		\textbf{network flows}  (e.g., source and destination IP addresses, amount of data sent over the network, duration of connections, 	time-based communication patterns);
		\textbf{specific applications}  (e.g., features derived from Microsoft Office Suite or other applications typical for an insider).

	\subsubsection*{\textbf{Capability.}} The capability features represent ``the demonstrated skill level of an insider monitored by the IT system.''
	Examples include:
		\textbf{consumption of system resources}  (e.g., consumption of CPU and RAM by the user across different sessions -- the higher the RAM and CPU usage, the higher the insider's level of sophistication may be);
		\textbf{application usage} (e.g., unique applications run by an insider in different sessions -- the higher the number of unique applications, the higher the insider's sophistication may be; multiple applications simultaneously run by the insider across different sessions -- the higher is the number of applications simultaneously run per session, the more sophisticated the insider may be).

\section{\textbf{Best Practices and Guidelines}}\label{appendix:best-practices}
Best practices and guidelines represent procedural defense countermeasures against insider threat which may, however, involve recommendations about the usage of technical controls. 
In this section we summarize different works in this area developed by academia, industry, and government.

\paragraph{\textbf{General}}
The major contribution to general best practices was made by CERT and the US Secret Service, both of which released several versions of \textit{common sense guides} (e.g.,~\cite{2016-Collins-CERTv5}) and in addition addressed specific areas such as government~\cite{2008-Kowalski}, the financial sector~\cite{2012-Cummings}, the critical infrastructure sector~\cite{2005-Keeney}, and the cloud environment~\cite{2012-Claycomb}.
Mitigation of malicious insiders from the cloud computing perspective is also described in~\cite{2011-Kandias}.
Technologies that can be used to control insider threat are addressed by Cole and Ring in chapter~9 of their book~\citeyear{2006-Cole}, while other countermeasures aimed at survivability are present in chapter~10. 
In addition to technical controls, a wide spectrum of procedural best practices and countermeasures are discussed in~\cite{1999-Hayden},~\cite{1999-Anderson}, and~\cite{2010-Sarkar}. 
Based on experiences with different organizations, Guido and Brooks~\citeyear{2013-Guido} described the components needed for an insider threat mitigation and auditing program and discussed several best practices. 
A wide range of factors influencing unintentional insider threat, along with best practices for its mitigation were discussed by Greitzer et al.~\citeyear{2014-Greitzer}.
Similarly aimed at unintentional data breaches, Liginlal et al.~\citeyear{2009-Liginlal} proposed three strategies for managing human error: \textit{avoidance}, \textit{interception}, and \textit{correction of error}, which were projected onto various data representations. 
McCormick~\citeyear{2008-Mccormick} proposed a number of ways for improving both administrative and technical controls for data leakage in a typical enterprise, while highlighting enterprise DLP programs. 
Data leakage was also addressed by Miller and Maxim~\citeyear{2015-Miller} who highlighted identity and access management along with DLP.

\paragraph{\textbf{Management and High Level View}}
This subsection includes papers dealing with insider threat from the managerial perspective, often presenting a high level view of addressing this problem. 
Steele and Wargo~\citeyear{2007-Steele} provided guidelines for the prevention and mitigation of insider threats. 
Their guidelines include cyclic stages of assessment, prioritization \& review, and remediation; all of these are governed by administrative methods (policies, procedures, human resources, awareness training, and education). 
Blackwell~\citeyear{2009-Blackwell} discussed aspects of a security model for the investigation and evaluation of organizational security in three layers: the \textit{social/organizational layer}, \textit{logical layer} (i.e., intangible entities), and \textit{physical layer} (i.e., tangible entities). 
In a similar vein, Viduto et al.~\citeyear{2010-Viduto} proposed a generic onion skin model for hardening the defense against malicious insiders, which consists of four layers (physical, technical, logical, and assets), each of which contains specific countermeasures. 
Gritzalis et al.~\citeyear{2014-Gritzalis} proposed mitigating strategies at the business process level by: 1) designing secure business processes, 2) performing risk assessment, and 3) monitoring each business process, while inferring conclusions.
Coles-Kemp and Theoharidou~\citeyear{2010-Coles-Kemp} examined  information security management practices with regard to the insider threat and further elaborated on crime theories that provide methods for information security management design. 
Colwill~\citeyear{2009-Colwill} examined many human factors (including technical, social, business, economic, and cultural factors) that can be used for assessing insider threat.

\paragraph{\textbf{Monitoring and Analysis of Incidents}}
Murphy et al.~\citeyear{2012-Murphy} proposed best practices and recommendations for the design of detection methods, as well as for the analysts of insider threat incidents working with them. 
In a similar vein, Gheyas and Abdallah~\citeyear{2016-Gheyas} identified several best practices related to the design of detection methods. 
Inspired by the overwhelming amount of monitoring alerts, Colombe and Stephens~\citeyear{2004-Colombe} proposed unsupervised anomaly-based visualization of RealSecure alerts, which enables the analyst to distinguish between non-automated and automated insider activity and pay attention to more interesting alerts. 
Doss and Tejay~\citeyear{2009-Doss} developed a model describing how analysts can use existing security tools, such as anti-virus, intrusion detection systems, and log analysis tools, to detect insider attacks.

\paragraph{\textbf{Standards}}
Insider threat has also been addressed by several standards. 
For instance, ISO 17799 provides a set of recommendations for information security management, and more specifically, the \textit{personnel security} category mentions controls aimed at the protection of an IS from accidental and deliberate insider threats~\cite{2005-Theoharidou}. 
Other related standards are ISO 27001 and ISO 27002, in which \textit{control domains} and \textit{control areas} address this topic~\cite{2010-Coles-Kemp,humphreys2008information}. 
According to Coles-Kemp and Theoharidou~\citeyear{2010-Coles-Kemp}, three distinct categories of controls can be identified in ISO 27002: ``controls identifying insiders from outsiders, controls used to identify unexpected insider behavior, and controls used to influence the development of an organization's security culture.''

\section{\textbf{Decoy-Based Solutions}}\label{appendix:decoys}
According to Spitzner~\citeyear{spitzner2003honeypots}, ``a honeypot is a security resource whose value lies in being probed, attacked, or compromised.''
Therefore, a honeypot has no production value, and any activity performed with it is suspect by nature. 
Although the primary purpose of this concept is the detection of external threats, the honeypot can also be utilized in the area of insider threat detection. 
A~honeytoken is an example of a decoy-based concept that provides fake information delivered by an internal legitimate system (e.g., record in database, files in a repository).

One of the first works in this area is~\cite{2003-Spitzner}, where the author discussed a few options for the detection of spies by placing false emails in mailboxes, planting honeytokens within search engine results, and \textit{``enriching"} network traffic with spurious data. 
Honeytokens as websites decoying insiders were used by Maybury et al.~\citeyear{2005-Maybury}.
Decoys for malicious insiders were also addressed by Bowen et al.~\citeyear{2009-Bowen-3} who proposed using trap-based documents with bogus credentials, as well as stealthily embedded beacons that signal an alert when the document is opened. 
Detection of masqueraders perpetrating data exfiltration in the Windows XP environment was addressed by Salem and Stolfo~\citeyear{2011-Salem} who designed a decoy document access sensor that compares a keyed-hash message authentication code (HMAC) embedded in each decoy document with HMAC computed over a document loaded into memory.
Virvilis et al.~\citeyear{2014-Virvilis} proposed several deceiving techniques for insiders and APTs, which were divided into two categories: 1) \textit{attack preparation} (e.g., fake entries in robots.txt, passwords in HTML comments, invisible HTML links, fake DNS records, and fake social network avatars), and 2) \textit{exploitation \& exfiltration} (e.g., honeytokens in databases, decoy files, and accounts). 
As part of their reconnaissance deception system, Achleitner et al.~\citeyear{2016-Achleitner} utilize honeypots to detect insider adversaries such as APTs performing reconnaissance.

\section{\textbf{Other Practical Solutions}}\label{appendix:other-solutions}
This section provides examples of other practical tools that can be applicable for the  detection or forensic investigation of insider threat incidents. 
In the majority of the cases, the authors of these tools do not provide detailed implementation descriptions, as these tools are usually commercial products.
The examples included here are grouped into three categories: \textit{detection}, \textit{SIEM}, and \textit{audit tools}. 
The detection category represents products that use several data sources to raise alarms for suspicious or anomalous activities. 
Security information and event management (SIEM) systems serve as log aggregation engines that provide consolidated views of the disparate events and alerts they analyze.
Audit tools can serve forensic and audit purposes, which may potentially lead to the detection of suspicious activities attributed to insider threat.
Note that this section does not contain references from our input literature database used for categorization, but rather it provides an overview of practical solutions that can be used for defense against insider threat.

\paragraph{\textbf{Detection}}
The examples of tools that we include in this category can be divided into two groups: \textit{misuse-based} tools, and \textit{machine learning-based} tools. 
Misuse-based tools are based on rule matching and include Raytheon's SureView~\citeyear{2015-SureView-Url}, the Ekran System~\citeyear{0000-EkranSystem-Url}, and SentinelOne~\citeyear{0000-sentinelone}. 
ObserveIT~\citeyear{0000-ObserveIt-Url} utilizes a different solution that involves a record of screen activity that may help in providing evidence in the case of an insider threat incident. 
In addition, there have been efforts made and incentives have been used to deploy machine learning techniques in commercial environments. 
Darktrace~\citeyear{0000-DarkTrace-URL} is based on advanced unsupervised machine learning techniques that utilize Bayesian probabilistic approaches to assess users' behavior, an entity, and software. 
Veriato 360~\citeyear{0000-Veriato360-Url} and Veriato Recon~\citeyear{0000-VeriatoRecon-Url} (from Veriato, formerly known as SpectorSoft) combine various machine learning techniques in order to detect anomalies in users' behavior.
Moreover, this tool integrates psychological profiling of users, which is obtained through the users' language. 
Securonix~\citeyear{0000-securonix} provides a solution based on machine learning techniques for the detection of anomalous behavior in a group of users, referred to as a control group. 
This tool is particularly aimed at the detection of data exfiltration.

\paragraph{\textbf{SIEM}}
Security information and event management (SIEM)~\cite{2010-Miller-SIEM} can provide statistical summaries of events over time, which can be aggregated and reported using various dimensions. 
We describe some examples of commercial and open-source tools, starting with a commercial tool, called ArcSight Express~\cite{0000-ArcSightExpress-Url} that provides a user tracking system based on events that can be correlated in real-time, and in the case of suspicious activity, an alert can be raised. 
Carbon Black's endpoint security platform~\citeyear{0000-CarbonBlack-Url} is capable of performing a full spectrum analysis that includes behavioral, reputation, machine learning, and SIEM analysis that could also be used for behavioral alerts regarding insider threat.
Kibana~\citeyear{0000-kibana} is an open-source SIEM based on the Elastic search engine  that has several features, such as geographical activity representation and time series of alarms. 
In addition, thanks to the nature of this tool and the availability of its source code, it is easy to build new custom components.

\paragraph{\textbf{Audit Tools}}
LUARM, an audit engine for insider IT misuse detection, was proposed by Magklaras et al.~\citeyear{2011-Magklaras}, and it logs actions associated with file system access and process execution, as well as actions at the network endpoint level. 
In the same vein, other commercial audit solutions propose similar features, such as MKinsight~\citeyear{0000-mkinsight} and ADAudit~Plus~\cite{0000-manageengine}, both of which are fully compatible with Windows enterprise deployments.

\section{Detection and Threat Assessment Approaches: Additional Works}\label{appendix:detection-extra}
In addition to the detection and threat assessment approaches discussed in Section~\ref{subsec:detection-and-assessment}, in this section we briefly describe the remaining papers of our input literature database which fit this subcategory, maintaining the categorization structure used in the main text of this survey.

\subsection{Conceptual Works}

\vspace{4mm}
\textbf{Anomaly-Based.}
Dealing with \textbf{role-based monitoring}, Phyo et al.~\citeyear{2004-Phyo-2} focused on insiders that misuse their privileges and proposed a detection concept for separation of duties violations; this concept involves matching input to a database of role-assigned actions. 
Incorporating attribute-based access control (ABAC), Bishop et al.~\citeyear{2009-Bishop-2} proposed an insider threat assessment concept for prioritizing users according to their access to resources, while considering \textbf{psychological indicators} gleaned from logs, HR records, and language affectation used in emails, IMs, and blogs. 
Greitzer and Hohimer~\citeyear{2011-Greitzer} proposed CHAMPION, a model-based belief propagation conceptual framework for reasoning about insider threat, which utilizes auto-associative memories and integrates psychological and cyber observables. 
In later work, Greitzer et al.~\citeyear{2012-Greitzer} proposed a risk assessment concept based on 12 behavioral and psychosocial indicators that serve as input to a Bayesian network, linear regression, and an artificial neural network (ANN).
Sasaki~\citeyear{2012-Sasaki} proposed a concept that creates a stimulating event that is used to prompt malicious insiders to behave anomalously, and consequently monitors the users' reactions. 
Berk et al.~\citeyear{2012-Berk} proposed BANDIT, a system that assesses the three components of the MOC model with respect to the user's baseline as well as to group behavior, producing two scores by calculating the Euclidean distance. 
In terms of \textbf{assessing the environment}, Chinchani et al.~\citeyear{2010-Chinchani} presented a modeling approach for insider threat assessment based on their previously proposed key challenge graph (KCG), in order to address the problem of finding an attack with minimal cost. 
The KCG concept was also applied by Ha et al.~\citeyear{2007-Ha} in their insider threat assessment graphical tool that displays possible attack trails. 
Naghmouchi et al.~\citeyear{2016-Naghmouchi} designed risk assessment graphs capturing the topological information of the network, including the assets and vulnerabilities associated with each asset, as well as the way these elements vary over time.

\subsection{Operational Works}

\vspace{4mm}
\textbf{Misuse-Based.}
Roberts et al.~\citeyear{2016-Roberts} designed a BN aimed at data exfiltration, consisting of four binary detectors: the presence of connected devices, after-hours work, visiting file sharing websites, and information about employees' dismissal. 
Graph-based misuse detection of malicious user interactions within \textit{Web applications} was proposed by Jaballah and Kheir~\citeyear{2016-Jaballah} who designed a gray box approach based on the subgraph isomorphic matching.

\vspace{4mm}
\textbf{Anomaly-Based.}
The bulk of the operational studies dealing with anomaly detection of insiders were conducted on \textbf{Unix/Linux command histories} and aimed at the masquerader attacker's model. 
Coull et al.~\citeyear{2003-Coull} proposed a masquerader detection approach that uses pair-wise sequence alignment to represent similarity between any two aligned sequences of commands in SEA. 
Wang and Stolfo~\citeyear{2003-Wang} utilized one-class Na\"{i}ve Bayes and one-class SVM for masquerader detection in SEA, comparing multivariate Bernoulli (binary) and multinominal (bag-of-words) approaches for feature representation. 
Yu and Graham~\citeyear{2006-Yu} employed a concept of finite state machine and a fuzzy inference system for modeling users; evaluation was performed on the PU dataset, as well as their custom dataset collected from 44 users. 
Szymanski and Zhang~\citeyear{2004-Szymanski} used one-class SVM and recursive mining that recursively searches for frequent patterns in a string of commands in SEA, and then they encoded such patterns with unique symbols and rewrote the string using this new coding. 
Oka et al.~\citeyear{2004-Oka-2} employed Eigen co-occurrence matrices and layered networks for the detection of masqueraders in SEA.
Latendresse~\citeyear{2005-Latendresse} employed a customized version of the Sequitur algorithm to generate context-free grammar from sequences of commands executed by a user; the author evaluated the approach on SEA. 
Similarly, dealing with the compression of commands in SEA, Posadas et al.~\citeyear{2006-Posadas} proposed a local hybrid method that combines HMM and session folding accomplished by the extraction of the user's context-free grammar. 
In order to deal with concept drift in the masquerader detection problem, Sen~\citeyear{2014-Sen} proposed applying instance weighting as an updating scheme for the Na\"{i}ve Bayes model; she evaluated the approach on SEA. 
Considering \textbf{file system logs} as a data source, Cami{\~n}a et al.~\citeyear{2014-Camina}  proposed detection systems for masqueraders utilizing Markov chains and Na\"{i}ve Bayes as one-class techniques that were evaluated on the WUIL dataset. 
Gates et al.~\citeyear{2014-Gates} proposed detection of information theft by comparing input with a history of the user's file access (and that of his/her peers) by similarity functions working over hierarchical file system structure; evaluation was performed on logs from a source code repository containing injected attackers. 
Aimed at per process file path diversity assessment, Wang et al.~\citeyear{2014-Wang} proposed a system for detection of masqueraders searching for sensitive files in Windows machines. 
Dealing with \textbf{system call sequences}, Nguyen et al.~\citeyear{2003-Nguyen} proposed a system for the detection of insiders in Unix systems based on the profiling of: 1) file system access executed by system users, and 2) human and system processes having either a fixed number of children or accessed files. 
The evaluation was performed on detection of insiders exploiting local buffer overflow vulnerabilities.
The detection of insiders in system calls was also addressed by Parveen et al.~\citeyear{2011-Parveen} who employed an ensemble of one-class SVMs as a stream-based approach that coped with the concept drift problem.
In contrast to the previous papers, we now present several examples dealing with \textbf{general cyber observables}. 
Shavlik and Shavlik~\citeyear{2004-Shavlik} proposed a masquerader detection method that creates statistical profiles of Windows 2000 users and intakes over 200 features measured in seven historical variants, weighted using a custom algorithm called Winnow.
Raissi-Dehkordi and Carr~\citeyear{2011-Raissi-Dehkordi} utilized one-class SVM to analyze features derived from file/database sever access and network monitoring in order to address information theft involving colluding insiders. 
User identification was the focus of Song et al.~\citeyear{2013-Song} who compared a Gaussian mixture model, Parzen method, and one-class SVM (in a one-vs-all classification setting); these were further optimized by the Fisher criterion for feature selection and evaluated on benign data of the RUU dataset.
Park and Giordano~\citeyear{2006-Park} proposed a combination of role-based and individual-based anomaly detection of information theft in the intelligence community.
Considering \textbf{psychosocial observables}, Greitzer et al.~\citeyear{2013-Greitzer-2} implemented a tool based on their previous work on personality traits~\citeyear{2012-Greitzer}, which performs word use analysis by LIWC with the Mahalanobis distance for the identification of outliers; they evaluated the approach on an email corpus containing injected outliers, and again later~\cite{2013-Brown} on the Enron dataset. 
Considering MOC indicators at an abstract level, Axelrad et al.~\citeyear{2013-Axelrad} developed a list of 83 indicators potentially related to insider threat (e.g., psychosocial and CWB indicators), which were further ranked and combined into a single score using a BN. 
Leveraging \textbf{NLP} techniques, Santos et al.~\citeyear{2008-Santos} presented intent-driven detection of malicious insiders spreading disinformation in intelligence reports by utilizing a BN that captures interests of analyst, context of knowledge, and changes of preferences over time.
Later, these authors extended their approach by a normalization procedure that divided an analyst's discrepancy value by his/her global correlation value~\cite{2012-Santos}.

\vspace{4mm}
\textbf{Classification-Based.}
Wu et al.~\citeyear{2009-Wu} proposed the detection of anomalous \textbf{database} transactions by two Na\"{i}ve Bayes classifiers modeling role profiles and user profiles, respectively, while utilizing six input features: user ID, role ID, time, IP address, access type, and SQL statement. 
Kim and Cha~\citeyear{2005-Kim} applied SVM with radial basis function for the detection of masqueraders in blocks of \textbf{Unix commands}, while they applied a simple voting scheme to assess a super-block of these commands.
Detection of masquerades in Unix commands was also addressed by Maxion~\citeyear{2003-Maxion}, who applied a Na\"{i}ve Bayes classifier with an updating scheme, which considered frequency of particular commands and their parameters for each user.
Yung~\citeyear{2004-Yung} extended a Na\"{i}ve Bayes classifier by self-consistency that kept both of the classes updated consistently with new data by the expectation-maximization algorithm.  
Killourhy and Maxion~\citeyear{2008-Killourhy} proposed the detection of super-masqueraders -- those issuing never-before-seen-commands (NBSC) -- in Unix command sequences using an enhanced Na\"{i}ve Bayes classifier with a simple NBSC detector.
An approach to modeling user interaction with a \textbf{GUI-based application} was proposed by El Masri et al.~\citeyear{2014-ElMasri} who focused on masquerader detection and used ensemble-based classifiers, such as Random Forest and Ada-Boost with Random Forest, and evaluated the results on Microsoft Word commands of the MITRE OWL dataset. 
Considering \textbf{sentiment analysis} and \textbf{NLP}, Taylor et al.~\citeyear{2013-Taylor} proposed detection of insiders perpetrating data exfiltration by binary logistic regression using several LIWC categories, together with linguistic style matching in email messages.

\vspace{4mm}
\textbf{Unsupervised Detection of Outliers.}
In the general \textbf{cyber observables} vein, Young et al.~\citeyear{2014-Young} evaluated previously designed methods~\cite{2013-Senator} as an ensemble against individual methods and scenario-based outlier detectors.
Investigating various activity domains, Eldardiry et al.~\citeyear{2013-Eldardiry} proposed detection of anomalies  representing insider threat by k-means clustering and Markov chain algorithms.
Dealing with the concept drift problem, Parveen et al.~\citeyear{2011-Parveen-1} combined a stream mining approach with an ensemble of GBADs for the detection of insiders in \textbf{system call} activities.
Wurzenberger and Kastner~\citeyear{2016-Wurzenberger} proposed a scalable approach for the detection of data exfiltration in \textbf{databases} by applying high performance bioinformatics clustering tools. 
Considering aspects of \textbf{mutual interactions of users}, data exfiltration in collaborative information systems (CIS) was addressed by Chen et al.~\citeyear{2012-Chen} who utilized nearest neighbor networks applied onto access logs of a healthcare CIS, observing that normal users tend to form communities unlike illicit insiders.

\section{Additional Categorizations of Operational Approaches of the Detection and Threat Assessment Subcategory}
\label{appendix:supplementary-Material}
In addition to the main categorization that is based on intrusion detection and machine learning, we propose the use of two other categorizations that can also be applied on operational works dealing with the detection and assessment of insider threat: 1) categorization based on \textit{\textbf{the dataset setting used for evaluation}}, and 2) categorization based on \textit{\textbf{the feature domains}}.
The former distinguishes among six types of selected dataset settings used in experiments, which include detection of \textit{masquerader attacks}, \textit{traitor attacks}, \textit{miscellaneous malicious attacks}, \textit{substituted masquerader attacks}, \textit{unintentional insider incidents}, and \textit{identification of users}. 
The majority of these settings are in accordance with the dataset categorization introduced in Section~\ref{Sec:Datasets}, however we emphasize that various types of datasets can be utilized in different types of settings; for example, the masquerader-based datasets can be utilized in the user identification task (e.g.,~\cite{2013-Song}), and the identification/authentication-based datasets can be utilized in the detection of substituted masqueraders (e.g.,~\cite{2003-Maxion,2014-ElMasri,2006-Yu}).
The latter type of categorization considers a taxonomy of detection feature domains~\cite{2016-Gheyas} based on the MOC model. 
For details about this taxonomy, we refer the reader to Appendix~\ref{appendix:taxonomy-features}. 
The three types of categorization (the main one and two additional ones) are applied to the operational works in Table~\ref{tab:detection-by-dataset-and-features} and~Table~\ref{tab:detection-by-dataset-and-features-2}.
In comparison to the original categorization of feature domains~\cite{2016-Gheyas}, several new feature domains were added to the tables in order to reflect all of the operational works of our literature database.
Note that some of the approaches listed in the tables contain a particular source of data in the description, however the feature domain referring to that data source is not marked (or vice versa). 
This is because some approaches apply non-data-inherent feature domains for data analysis.
For example,~\cite{2013-Brown,2013-Greitzer-2} use emails as a data source, however their approach works only with the text information of those emails and considers motive-based feature domains (i.e., personality factors, current emotional states); this is in contrast with~\cite{2015-Legg}, whose approach also analyzes recipients and senders of emails.

\begin{table}            
	\centering{}
	\smallskip
	\begin{centering}    
		\includegraphics[width=0.97\textheight, angle=-90]{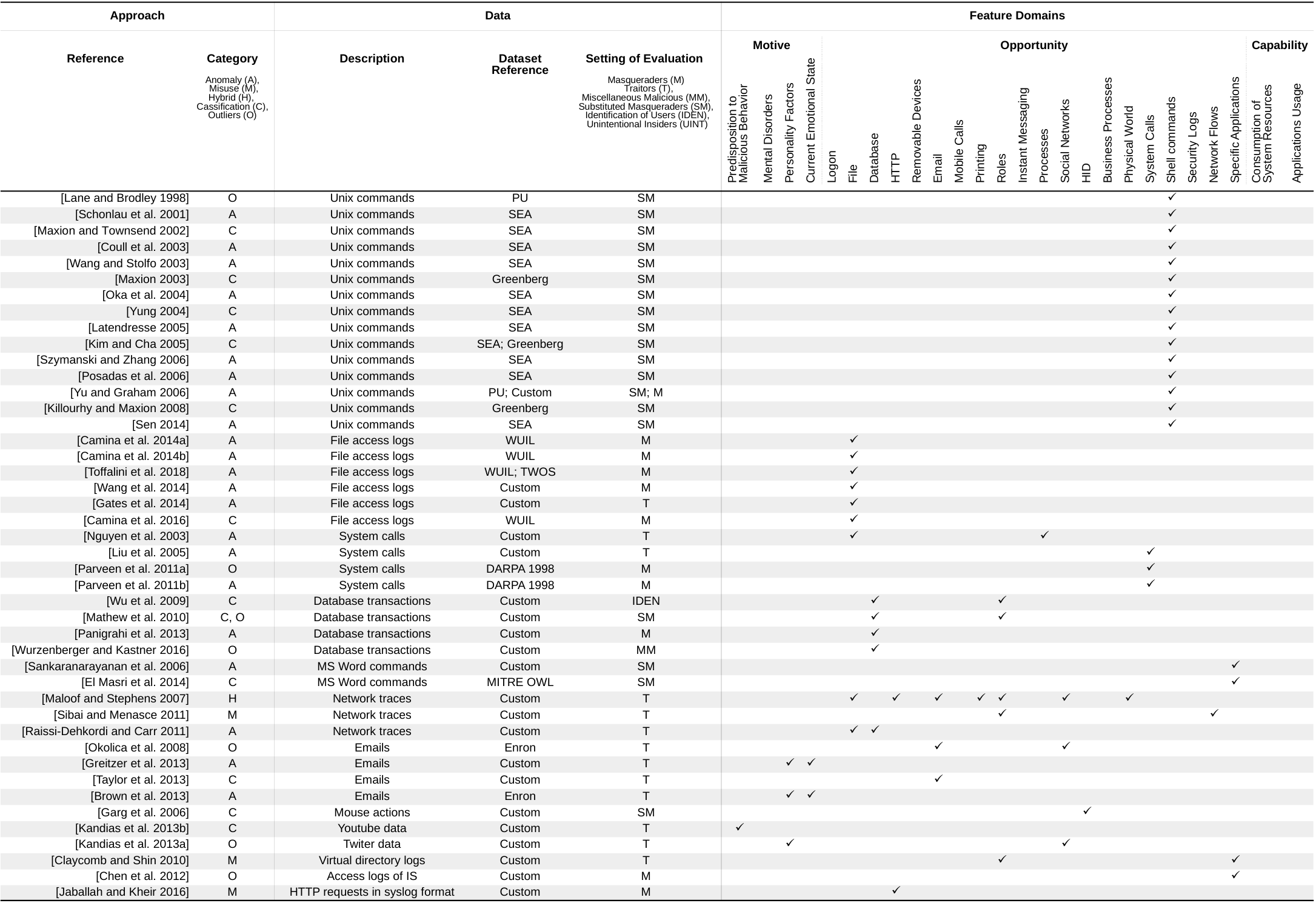}                
	\end{centering}
	\vspace{0.2cm}
	\caption{Categorization of the operational approaches (part 1/2)}
	\label{tab:detection-by-dataset-and-features}		 
\end{table}  

\begin{table}            
	\centering{}
	\smallskip
	\begin{centering}    
		\includegraphics[width=0.97\textheight, angle=-90]{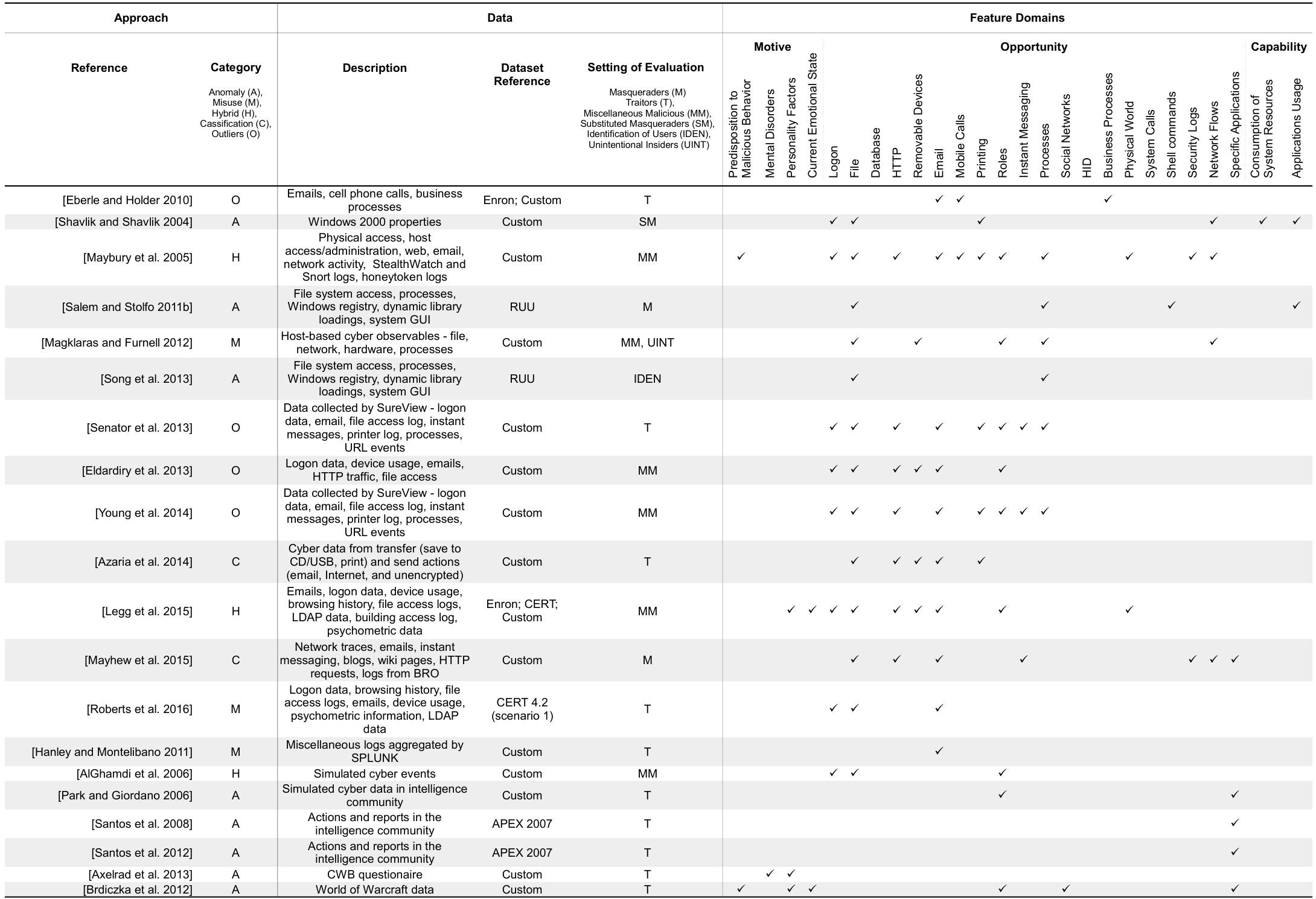}                
	\end{centering}
	\vspace{0.2cm}
	\caption{Categorization of the operational approaches (part 2/2)}
	\label{tab:detection-by-dataset-and-features-2}		 
\end{table}

\clearpage
\section{All Research Contributions in Insider Threat Field}\label{appendix:overal-categorization}
\begin{figure*}[!h]
	\vspace{-0.2cm}
	\centering
	\includegraphics[width=0.91\textheight,angle=-90,origin=c]{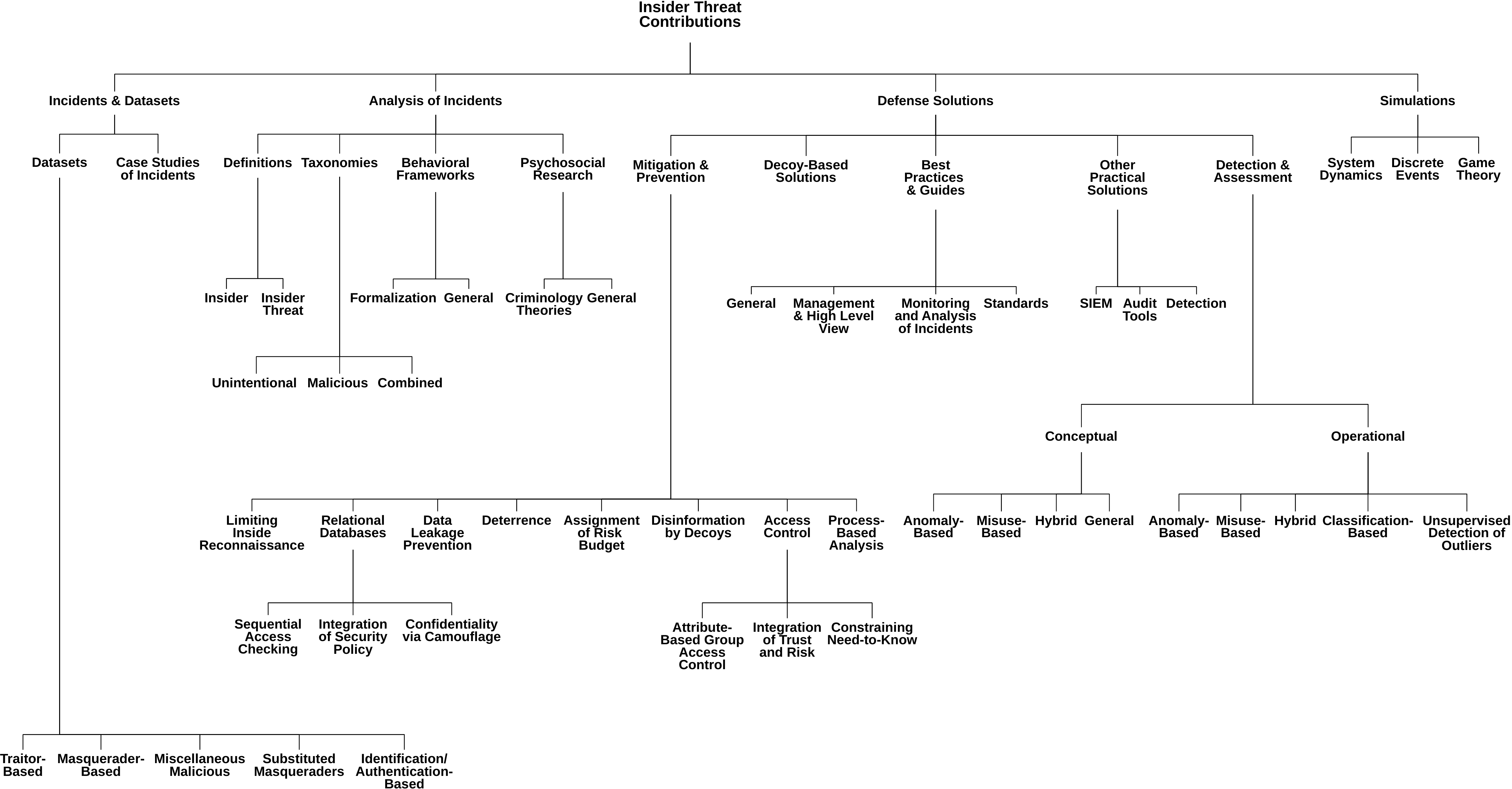}
	
	\vspace{-0.35cm}
	\caption{Categorization of research contributions in insider threat field \label{appendix:categorization-of-contributions}}
	\vspace{-0.4cm}
\end{figure*}